\RequirePackage[T1]{fontenc}
\documentclass[12pt]{article}

\usepackage[height=8.85in,width=6.45in]{geometry}

\usepackage[utf8]{inputenc}
\usepackage{amsmath}
\usepackage{amssymb}
\usepackage{mathtools}
\usepackage{physics}
\numberwithin{equation}{section}
\usepackage{slashed}
\usepackage{braket}
\usepackage[svgnames]{xcolor}
\usepackage{soul}
\usepackage[nolist,printonlyused]{acronym}
\def\acroswithtooltips{1}
\if\acroswithtooltips1
\usepackage{pdfcomment}
\newcommand{\tooltipacro}[2]{\acro{#1}[\pdftooltip{#1}{#2}]{#2}}
\fi

\hypersetup{colorlinks,citecolor=DarkGreen,linkcolor=FireBrick,linktocpage}

\usepackage{varioref}
\usepackage[nameinlink]{cleveref}

\crefname{table}{table}{tables}
\Crefname{table}{Table}{Tables}
\crefname{figure}{figure}{figures}
\Crefname{figure}{Figure}{Figures}

\usepackage{cite}
\usepackage{graphicx}
\usepackage{tikz}
\usepackage{tikz-cd}
\usepackage{times}
\usepackage{courier}
\usepackage{bm}
\usepackage{subfig}
\usepackage{ascmac}
\usepackage{tikzsymbols}
\usepackage{tensor}

\usepackage{mdframed}

\makeatletter
\AddToHook{cmd/appendix/before}{\def\cref@section@alias{appendix}}
\makeatother

\renewenvironment{figure}[1][]{
  \begin{originalfigure}[#1]
    \begin{mdframed}[linecolor=black!0,backgroundcolor=black!1]
}{
    \end{mdframed}
  \end{originalfigure}
}
\renewenvironment{table}[1][]{
  \begin{originaltable}[#1]
    \begin{mdframed}[linecolor=black!0,backgroundcolor=black!1]
}{
    \end{mdframed}
  \end{originaltable}
}

\newenvironment{eqaed}
    {\begin{equation}
    \begin{aligned}
    }
    { 
    \end{aligned}
    \end{equation}
    \ignorespacesafterend
    }

\usepackage{tikz}
\usepackage{dsfont}
\usetikzlibrary{calc}

\begin{document}

\begin{titlepage}

\begin{flushright}
MPP-2026-102
\end{flushright}

\vskip 3cm

\renewcommand*{\thefootnote}{\fnsymbol{footnote}}

\begin{center}

{\Large \bfseries String dualities and wedge singularities}

\vskip 1cm
Ivano Basile$^{1,2,}$\footnote{\url{ibasile@mpp.mpg.de}}, Dieter L\"{u}st$^{1,3,}$\footnote{\url{luest@mpp.mpg.de}}
\vskip 1cm

\begin{tabular}{ll}
$^1$ & \emph{Max-Planck-Institut f\"{u}r Physik (Werner-Heisenberg-Institut)}\\
& \emph{Boltzmannstraße 8, 85748 Garching, Germany} \\
$^2$ & \emph{INFN sezione di Torino}\\
& \emph{via Pietro Giuria 1, 10125 Torino, Italy} \\
$^3$ & \emph{Arnold-Sommerfeld Center for Theoretical Physics, Ludwig Maximilians Universit\"at M\"unchen}\\ & \emph{The\-re\-sienstraße 37, 80333 M\"unchen, Germany}
\end{tabular}

\end{center}

\begin{abstract}
    \noindent We study strings propagating in backgrounds with a wedge singularity, namely whose internal sector describes a wedge sum of closed manifolds. We focus on the wedge sum of two circles, which was recently argued to provide a quantum geometry for an M-theoretic description of type 0 strings, along with a much wider non-supersymmetric duality web stemming from quotients thereof. In the context of this proposal, we investigate whether the worldsheet features of type IIA strings, together with strong-coupling ingredients, can consistently reproduce the expectation of a weakly coupled type 0A frame. To this end, we combine worldsheet effects due to the wedge singularity with the emergence proposal applied to D0-branes probing it. We find that the resulting potential reproduces the correct tree-level mass of the tachyon in a specific scaling limit. We also discuss the possibility of kinematic obstructions to our worldsheet approach using the framework of topological modular forms, and comment on some puzzles and open questions.
\end{abstract}

\end{titlepage}

\setcounter{tocdepth}{2}
\tableofcontents

\renewcommand*{\thefootnote}{\arabic{footnote}}
\setcounter{footnote}{0}


\section{Introduction}\label{sec:introduction}

Recently, it was proposed that the duality web of ten-dimensional perturbative string limits can be completed by considering M-theory backgrounds whose internal sector contains the singular space $S^1 \vee S^1$ \cite{Baykara:2026gem, Altavista:2026evd, Baykara:2026vdc, Altavista:2026brr}. Its simplest instance is the proposal that type 0A string theory in ten dimensions arise as a reduction of M-theory on this space, whose rules are yet to be defined to a complete extent. Some salient features of this identification are the following:

\begin{itemize}
    \item \textbf{Quantum geometry.} The singular space $S^1 \vee S^1$ departs slightly from the category of manifolds, yet it is not completely abstract or emergent such as what one sees in stringy matrix models (BFSS, IKKT, DVV, ...) or AdS/CFT (and holography in general). As a result, physically there appears to be no meaningful sense in which this space can be made large (in eleven-dimensional Planck units). To corroborate this view, the one-loop potential\footnote{Strictly speaking, the potential considered in \cite{Baykara:2026gem} corresponds to the definition of the Hamiltonian for unstable systems which gives finite, but non-real, eigenvalues, rather than real but \ac{IR}-divergent values.} of the type 0A string seems to provide an energy barrier to strong coupling.
    \item \textbf{Identification of the dilaton and tachyon.} Relatedly, and analogously to the case of type IIA strings, the dilaton is related to the overall size of this singular space. The novelty of this setup is that the tachyon is also geometrized, now in the difference between the two sizes.
    \item \textbf{Search for a complete framework.} The peculiar ``in-between'' character of this proposal makes it hard to place within a firm theoretical setting, where the mathematical rules and their physical interpretations are clear. Hence, research in this direction has thus far proceeded by analogy with scenarios that are better understood. As a result, only some aspects of the story can be explained, and they hinge on several seemingly \emph{ad hoc} choices; most prominently, whether fields obey the \ac{CRP} or the \ac{DRP} \cite{Baykara:2026gem}. In general, what is allowed and what is not remains mostly obscure. Still, it is rather tantalizing that a simple recipe can reproduce correctly more physical data than the mere input would suggest \cite{Altavista:2026brr}.
\end{itemize}

\noindent This state of affairs motivates a more grounded approach to further consistency checks. Since perturbative string theory has a rich history of resolving in a consistent fashion settings which do not make sense in effective field theory, the basic idea is to seek a worldsheet description of wedge singularities. Since in the proposal of \cite{Baykara:2026gem} the singular space $S^1 \vee S^1$ plays a distinguished non-perturbative role replacing the usual M-theoretic extra dimension, one must exercise care in comparing this proposal to perturbative strings on the same space. At best one can expect that a relation between the two, if any, be provided by a duality action; this is because the dilaton is never geometrized in perturbative string theory. Thus, our tentative approach is to further reduce M-theory on a standard circle (with periodic spin structure) and hypothesize that the order of reductions can be permuted. This presumably results in type IIA strings on $S^1 \vee S^1$, which---if at all consistent---would occupy a different duality frame than type 0A strings on a circle. The link between the two frames would then likely be non-perturbative, due to the different (quantum-)geometric origin of the dilaton. Another hint in this respect is that, if type IIA strings allow such wedge singularities, there would be no obstruction to making the singular space arbitrarily large (in string units), at least at tree level. Perhaps one-loop effects would already show some obstruction, akin those identified in \cite{Baykara:2026gem, Baykara:2026vdc}; however, as we shall see, connecting the singular geometry to a weakly coupled type 0 dual frame seems to require strong string coupling in the type II frame, possibly jeopardizing such an approach in this case. \\

\noindent As we will discuss, it seems likely that a physically sensible worldsheet formulation of wedge singularities be incompatible with (unbroken) spacetime supersymmetry. Still, some amount of worldsheet supersymmetry may be required to account for the doubling of \ac{RR} vacua; this is automatically the case if we work with type IIA strings, but the physical consequences of wedge singularities in the fermionic sectors are to be explored to check whether the overall picture is consistent. For the time being, in this work we focus on selected aspects of this putative duality, namely the type 0A tachyon, the doubling of \ac{RR} sectors and some topological obstructions to a worldsheet description of wegde singularities. We find a scaling limit in which the extrapolation of naive duality relations seems to be consistent with the picture of \cite{Baykara:2026gem}; in particular, the tree-level mass of the type 0A tachyon can be reproduced by the emergence proposal applied to type IIA D0-branes in a certain limit, thanks to non-trivial worldsheet effects. At the topological level we find no obstructions to a worldsheet description of $S^1 \vee S^1$ from the mathematics of topological modular forms, although other settings (in particular, even-dimensional wedge sums) are already obstructed at this level. \\

\noindent \paragraph{Summary.} The paper is organized as follows. In \cref{sec:dualities_setup} we provide an overview of the proposals of \cite{Baykara:2026gem, Altavista:2026evd, Baykara:2026vdc, Altavista:2026brr} regarding type 0A string theory, and we introduce the relevant scaling limit from the type IIA perspective. In particular, \cref{sec:K-theory} contains some arguments to the effect that type 0A \ac{RR} charges are doubled. In \cref{sec:strings_on_wedge_singularities} we introduce our tentative worldsheet description of wedge singularities, discussing in detail the stringy corrections to the geometry and the dilaton gradient that arise from $\alpha'$-effects. In \cref{sec:emergence} we apply the emergence proposal to the light species in the type IIA frame, arguing that a specific strong-coupling limit matches the tree-level mass of the type 0A tachyon. We also discuss other aspects of the potential and how stringy corrections affect the contributions of winding modes and instanton terms. In \cref{sec:cobordisms_singularities} we broaden the discussion to more general wedge singularities: in \cref{sec:more_general_wedge_singularities} we extend the worldsheet approach developed in \cref{sec:strings_on_wedge_singularities}, while in \cref{sec:worldsheet_cobordisms} we discuss cobordism obstructions from the perspective of the Stolz-Teichner conjecture. We provide some concluding thoughts in \cref{sec:conclusions}.

\section{Dualities and setup}\label{sec:dualities_setup}

The main setting we shall attempt to define and study is type IIA string theory\footnote{The specifics of type IIA will not enter our worldsheet description directly. Rather, they will only be relevant for the discussion of D-brane physics and the duality within the M-theoretic picture of type 0A and type IIA strings.} on $S^1 \vee S^1$. We will later extend some aspects of the discussion to more general wedge singularities. Following the proposals of \cite{Baykara:2026gem, Altavista:2026evd, Baykara:2026vdc, Altavista:2026brr}, and denoting by the symbol $\mathcal{T}/M$ the reduction of a theory $\mathcal{T}$ over a space $M$, we postulate the dualities
\begin{eqaed}\label{eq:duality_M-IIA-0b}
    \frac{\text{type IIA}}{S^1_{R_+} \vee S^1_{R_-}} = \frac{\text{M-theory}}{(S^1_{R_+} \vee S^1_{R_-}) \times S^1_{R_{11}}} = \frac{\text{type 0A}}{S^1_{R_{11}}} \, ,
\end{eqaed}
where $S^1_r$ denotes the circle of radius $r$ and we take $R_\pm = R(1 \pm T)$, with $T \ll 1$ the tachyon \ac{VEV} in the type 0A frame, and the respective string couplings $g_s$ and string scales $M_s$ are given by
\begin{eqaed}\label{eq:string_couplings}
    & g_s^{\text{IIA}} = (M_{11}R_{11})^{\frac{3}{2}} \, , \qquad M_s^\text{IIA} = M_{11} \, (g_s^\text{IIA})^\frac{1}{3} \, , \\
    & g_s^{\text{0A}} = (M_{11}R)^{\frac{3}{2}} \, , \qquad M_s^\text{0A} = M_{11} \, (g_s^\text{0A})^\frac{1}{3}
\end{eqaed}
in terms of the eleven-dimensional Planck scale $M_{11}$, up to positive numerical prefactors (which we shall set to unity in this work). Then, for small tachyon \acp{VEV} $T \ll 1$, one has\footnote{In \cite{Baykara:2026gem}, the complete form of \cref{eq:tachyon_VEV}, as well as an analogous formula for $g_s^\text{0A}$, were proposed by assuming a conjectural expression for the tension of D-branes as a function of the tachyon \ac{VEV} \cite{Garousi:1999fu}.}
\begin{eqaed}\label{eq:tachyon_VEV}
    T \overset{R_+ \to R_-}{\sim} \frac{R_+ - R_-}{R_+ + R_-} \, .
\end{eqaed}
In general, we take such parametric relations as a tentative starting point, given the speculative nature of these dualities and the absence of apparent supersymmetric protection mechanisms. It is then clear that ten-dimensional type 0A strings are recovered in the strong-coupling limit of the type IIA description. In order that this limit be consistent, the lightest states ought to match; the simplest limit is $g_s^\text{IIA} \to +\infty$, where D0-branes correspond to \ac{KK} species, resulting in ten-dimensional type 0A string theory according to \cref{eq:duality_M-IIA-0b}. \\

\noindent \textbf{A weakly coupled type 0A limit.} In order to probe the consistency of the proposed duality web more quantitatively, it would be useful to take a limit whose type 0A description be weakly coupled. Then, consistently with the \ac{ESC} \cite{Lee:2019wij}, an equi-dimensional limit requires that the light tower of D0-branes scale equally with the tower of excitations of the type 0A fundamental string which arises as $g_s^\text{0A} \to 0^+$. From \cref{eq:string_couplings}, one obtains
\begin{eqaed}\label{eq:string_coupling_map}
    g_s^\text{0A} = \frac{1}{(g_s^\text{IIA})^2} \, .
\end{eqaed}
This result also highlights that type 0B strings are recovered by T-duality. Indeed, according to the corresponding F-theoretic picture of \cite{Baykara:2026gem}, \cref{eq:string_coupling_map} implies that the string coupling $g_s^\text{0B}$ and the radius $R_\text{0B}$ in the type 0B frame be given by 
\begin{eqaed}\label{eq:0B_quantities}
    & g_s^\text{0B} = \frac{R}{R_{11}} \overset{\text{\cref{eq:string_coupling_map}}}{=} \frac{1}{(g_s^\text{IIA})^2} \overset{\text{\cref{eq:string_coupling_map}}}{=} g_s^\text{0A} \, , \\
    & M_s^\text{0B} R_\text{0B} = \frac{M_s^\text{0A}}{M_{11}^3 R_{11} R} \overset{\text{\cref{eq:string_coupling_map}}}{=} M_s^\text{0A}R_{11} = 1 \, .
\end{eqaed}
In the following we focus on this limit, but the ideas we shall present can be applied to, and tested against, other choices of scalings as well. \\

\noindent With these stipulations, according to \cref{eq:duality_M-IIA-0b,eq:string_couplings,eq:string_coupling_map}, the limit $g_s^\text{IIA} \to +\infty$ ought to recover weakly coupled type 0A strings in nine dimensions, whose tree-level physics is understood from the worldsheet approach and should thus be reproduced by strong-coupling effects in the type IIA frame. In order to discuss these effects, we shall first formulate a tentative worldsheet description of the latter, keeping in mind that any extrapolation to strong coupling is inevitably going to be speculative and qualitative at best.

\subsection{D-brane charge doubling: K-theory and graph spectra}\label{sec:K-theory}

Before moving on to the dynamics of our proposed worldsheet description, it is worth discussing some topological aspects of the story. In this sense, perhaps the most apparent qualitative difference between type II and type 0 strings is the doubling of \ac{RR} sectors, along with the corresponding D-brane content. This can be readily understood from the perturbative duality between type II and type 0 frames, where the additional \ac{RR} states arise in the twisted sector upon orbifolding by the fermion number. However, on account of the proposal of \cref{eq:duality_M-IIA-0b,eq:string_coupling_map}, this doubling should also be visible in the strong-coupling limit in the type IIA frame. This possibility raises an immediate puzzle: how could the worldsheet theory on $S^1 \vee S^1$ encode the fact that the different sectors ought to obey different resolution properties? In order to obtain a unique gravitational sector, including the B-field, naively it seems that the worldsheet description cannot involve the different resolutions separately. Rather, the wedge singularity has to be resolved non-perturbatively; this is indeed what we shall find in \cref{sec:emergence}. For the time being, we can approach the qualitative behavior of the wedge singularity in two ways: algebraic topology and discretization. \\

\noindent \paragraph{K-theory of bouquets.} We begin by discussing topological D-brane charges on wedge sums. In this section we specialize to $S^1 \vee S^1$, but the following considerations can be readily generalized, as in \cref{sec:more_general_wedge_singularities}. At the level of approximation we are working with, D-brane charges can be classified by topological K-theory \cite{Witten:1998cd,Horava:1998jy,Kaidi:2019tyf}; specifically, in type IIA strings compactified on a space $X$, the complex K-theory group $K^{10-p-\dim X}(X)$---more precisely, its reduced subgroup---encodes D$p$-brane charges\footnote{This is a special case of the general pattern with which ``freezing'' the internal space by dimensional reduction modifies topological charges \cite{Blumenhagen:2022bvh}.}, where the degree is shifted by one relative to the type IIB case \cite{Witten:1998cd,Horava:1998jy}. As pointed out in \cite{Witten:1998cd,Horava:1998jy}, this shift is reminiscent of the M-theory circle, which in the classification of type 0A charges from the type IIA frame would be replaced by $S^1 \vee S^1$. Although the bouquet $S^1 \vee S^1$ is not a manifold, we can still consider its (co)homology groups. This leads us to consider the complex K-theory groups $K^*(S^1 \vee S^1)$. The reduced subgroups $\widetilde{K}^*(S^1 \vee S^1) = \widetilde{K}^*(S^1) \oplus \widetilde{K}^*(S^1)$ follow from additivity under wedge sums, but in this case the conclusion follows directly from a simple instance of the cohomological Atiyah-Hirzebruch spectral sequence \cite{McCleary_2000}\footnote{In the spirit of \cite{Witten:1998cd,Blumenhagen:2022bvh}, perhaps it would be conceptually more appropriate to use the generalized cohomology theory represented by the suspended spectrum $\Sigma KU$ due to the degree shift in the type IIA frame. The computation can be carried out either way in general, and especially in this case due to Bott periodicity.},
\begin{eqaed}\label{eq:AHSS}
    H^{p}(S^1 \vee S^1, K^q(\text{pt})) \quad \Longrightarrow \quad K^{p+q}(S^1 \vee S^1) \, .
\end{eqaed}
If $X$ and $Y$ are connected, an argument employing the Mayer Vietoris sequence\footnote{In the case of $S^1 \vee S^1$, where only cohomology groups of degrees $p\leq 1$ are non-trivial, the same conclusion can be reached considering the free-product fundamental group $\pi_1(S^1 \vee S^1)=\mathbb{Z} * \mathbb{Z}$, whose abelianization yields the first cohomology group $H^1(S^1 \vee S^2,\mathbb{Z}) = \mathbb{Z}\oplus \mathbb{Z}$.} shows that $H^k(X \vee Y, \mathbb{Z}) = H^k(X, \mathbb{Z}) \oplus H^k(Y, \mathbb{Z})$ for $k>0$, while $H^0(X \vee Y, \mathbb{Z}) = \mathbb{Z}$. We have $K^0(\text{pt}) = \mathbb{Z}$ and $K^1(\text{pt}) = 0$, and thus the Atiyah-Hirzebruch spectral sequence (depicted in \cref{fig:ahss}) collapses at the $E_2$-page, since there cannot be non-trivial differentials. 

\begin{figure}[!ht]
\centering
\begin{tikzpicture}
\matrix (m) [matrix of math nodes,
    nodes in empty cells,
    nodes={
      minimum width=5ex,
      minimum height=5ex,
      anchor=center
    },
    column sep=1ex,
    row sep=1ex]{
      &      &      &      \\
  2 & \mathbb{Z} & \mathbb{Z} \oplus \mathbb{Z} & 0 \\
  1 & 0 & 0 & 0 \\
  0 & \mathbb{Z} & \mathbb{Z} \oplus \mathbb{Z} & 0 \\
     & 0 & 1 & 2 \\
};

\draw[thick]
  ($(m-2-1.east)!0.5!(m-2-2.west)+(0,3ex)$) --
  ($(m-5-1.east)!0.5!(m-5-2.west)+(0,-2ex)$);

\draw[thick]
  ($(m-4-1.south)!0.5!(m-5-1.north)+(-2ex,0)$) --
  ($(m-4-4.south)!0.5!(m-5-4.north)+(3ex,0)$);
\end{tikzpicture}
\caption{The $E_2$-page of the Atiyah-Hirzebruch spectral sequence for $K^*(S^1 \vee S^1)$. Entries $E_2^{p,q}$ with higher degree $p$ are on the right, and are trivial for $S^1 \vee S^1$. Entries with higher degree $q$ are higher on the page, and are 2-periodic due to Bott periodicity.}
\label{fig:ahss}
\end{figure}

\noindent All in all, we obtain
\begin{eqaed}\label{eq:K-groups}
    & K^\text{even}(S^1 \vee S^1) = \widetilde{K}^0(S^1 \vee S^1) \oplus \mathbb{Z} = \mathbb{Z} \, , \\
    & K^\text{odd}(S^1 \vee S^1) = \widetilde{K}^1(S^1 \vee S^1) = \mathbb{Z} \oplus \mathbb{Z} \, ,
\end{eqaed}
where we explicitly introduced the reduced subgroups. In particular, this result indicates that, according to the proposed duality web, topological charges pertaining to D$p$-branes with $p$ even experience a doubling, as befits a type 0A interpretation. From the M-theoretic description of the type 0A frame in \cref{eq:duality_M-IIA-0b}, where (to a crude degree of approximation) some topological charges are classified by integral (co)homology, these states arise wrapping M2-branes on $S^1 \vee S^1$; thus, the doubling in
\begin{eqaed}\label{eq:wedge_homology}
    H_1(S^1 \vee S^1, \mathbb{Z}) = \mathbb{Z} \oplus \mathbb{Z}
\end{eqaed}
is also suggestive of a correspondence between topological charges. Finally, for any space $X$, the above computation of reduced K-theory generalizes to
\begin{eqaed}\label{eq:general_K-theory}
    \widetilde{K}(X \wedge (S^1 \vee S^1)) = \widetilde{K}(\Sigma X \vee \Sigma X) = \widetilde{K}(\Sigma X) \oplus \widetilde{K}(\Sigma X) = \widetilde{K}^1(X) \oplus \widetilde{K}^1(X) \, , 
\end{eqaed}
which for $X = S^{9-p}$ (the one-point compactification of the normal space to a D$p$-brane) reduces to \cref{eq:K-groups}, consistently with doubling and the degree shift. \\

\begin{figure}[!ht]
    \centering
    \includegraphics[width=0.75\linewidth]{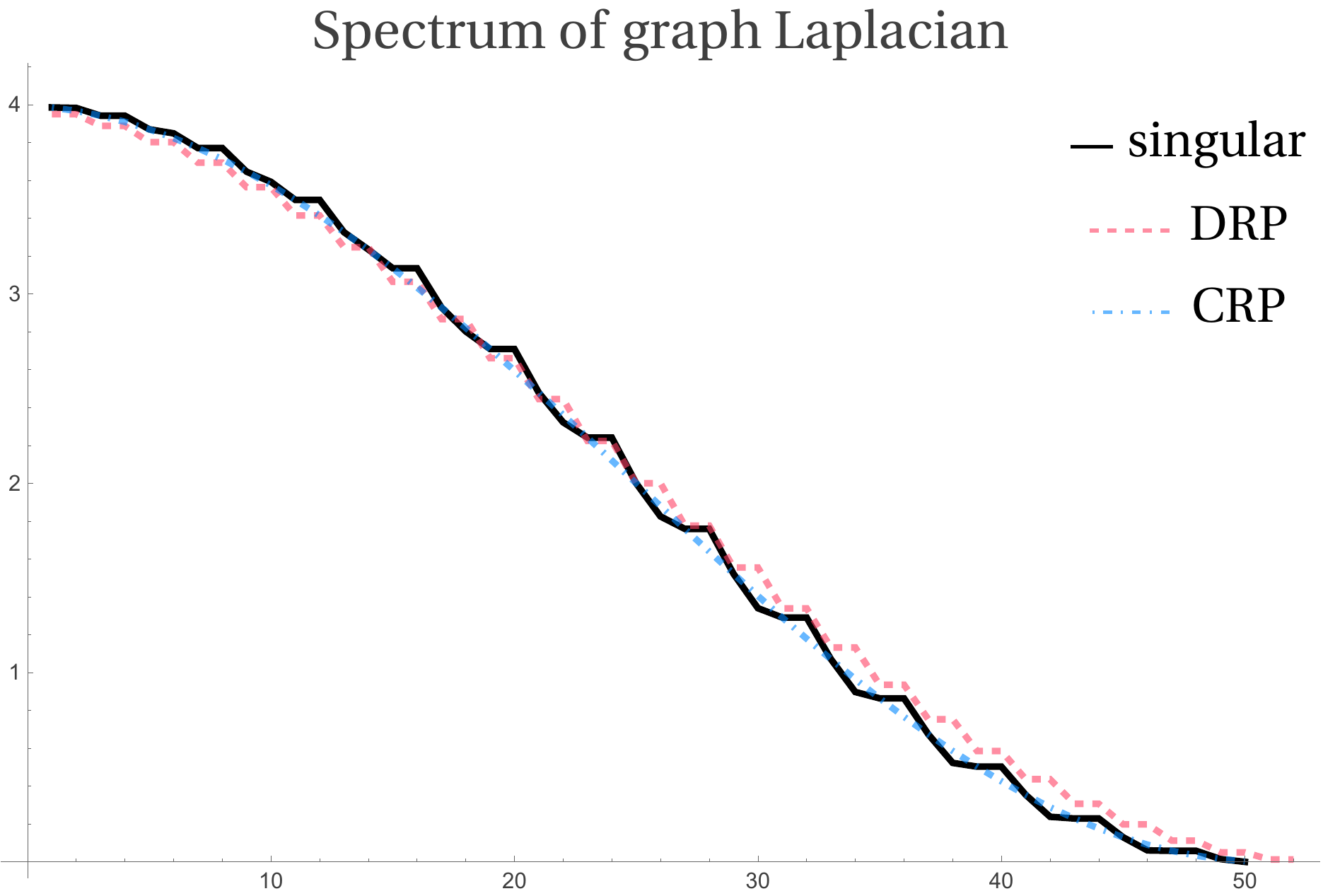}
    \caption{A plot of the spectrum of the Laplacian matrix associated with the graph discretization of $S^1 \vee S^1$ (black continuous line) with 50 nodes. The plot is accompanied by the Laplacian spectra corresponding to the graph discretizations of $S^1$ (\ac{CRP}, blue dot-dashed line) and $S^1 \sqcup S^1$ (\ac{DRP}, red dashed line). We observe that the regular plateaus in the \ac{DRP} spectrum, which encode the doubled degeneracy of \ac{KK} modes, are also present in the singular case, whereas they are absent in the familiar \ac{CRP} case.}
    \label{fig:spectrum_graph_laplacian}
\end{figure}

\noindent \paragraph{Laplacian spectrum on discretized bouquets.} Another approach to studying the wedge singularity of $S^1 \vee S^1$ (or other wedge sums) without an ambiguous resolution is discretization. Namely, one can replace $S^1 \vee S^1$ by a graph, and study the spectrum of its Laplacian (or Kirchoff) matrix in the limit of large numbers of nodes. Since D0-branes in the type 0A frame ought to correspond to \ac{KK} modes of $S^1 \vee S^1$ \cite{Baykara:2026gem}, this strategy may provide complementary hints on \ac{RR} doubling from the perspective of the putative strong-weak duality expressed by \cref{eq:duality_M-IIA-0b,eq:string_coupling_map}. The graph discretization $\Gamma_n$ of $S^1$ with $n$ nodes leads to a well-known circulant Laplacian matrix, whose spectrum
\begin{eqaed}\label{eq:S1_graph_spectrum}
    \lambda_k(\Gamma_n) = 4 \sin^2\frac{\pi k}{n} \overset{n \gg k}{\sim} \left(\frac{2\pi k}{n}\right)^{\!\!2}
\end{eqaed}
reproduces the correct scaling with a radius proportional to $n$. The case of $S^1 \sqcup S^1$, namely the disconnected resolution of $S^1 \vee S^1$, has the same spectrum with doubled degeneracy. In the same vein, we can approximate the behavior of would-be \ac{KK} modes on a wedge singularity by a discretized version, whose graph contains a node of degree four which connects to the two circles; therefore, its Laplacian matrix contains a doubled diagonal entry and some extra non-diagonal components of value $-1$. We computed the Laplacian spectrum of the graph discretization of $S^1 \vee S^1$ up to 50 nodes, finding similar doublings of degeneracy, albeit not constant as in the case of the disconnected resolutions. The results are plotted in \cref{fig:spectrum_graph_laplacian}, and indeed provide a further indication of \ac{RR} doubling for D0-branes in the putative type 0A dual frame. \\

\noindent Both of the above approaches crucially rely on the presence of the wedge singularity, which leads to the breaking of spacetime supersymmetry as we shall discuss in the following section. This is consistent with a putative dual type 0A picture, which we shall now attempt to formulate from the type IIA side via worldsheet methods.

\section{Strings probing the wedge sum of two circles}\label{sec:strings_on_wedge_singularities}

We now turn to the formulation and study of strings propagating on wedge singularities from the worldsheet point of view, focusing on the simplest but most relevant setting of $S^1 \vee S^1$. We postpone the discussion of more general wedge singularities to \cref{sec:more_general_wedge_singularities}. \\

\noindent Let us begin by attempting to describe (super)strings propagating on $S^1 \vee S^1$ from the worldsheet perspective. We shall focus on the internal sector and employ the \ac{RNS} formalism, although its specifics are not going to play a particularly relevant role over other formalisms. Since $S^1 \vee S^1$ is not a manifold, the usual \ac{NLSM} is not well-defined. As such, we need to work slightly harder to identify a physically reasonable alternative. To this end, our starting point is the framework introduced in \cite{Anastasi:2026cus, Altavista:2026edv, Tachikawa:2026jaj} to describe junctions of worldsheet \acp{CFT}. A possible alternative approach is coupling the direct sum of two compact bosons; however, it is not clear to us how to introduce such a coupling in a physically reasonable fashion: the direct sum of \acp{NLSM}, or equivalently the disjoint union of their target spaces, has no extrinsic notion of separation between the manifolds, and thus no notion of the possible resolutions of the singularity in their wedge sum. A seemingly more straightforward strategy is to appeal to the rich dynamics of \acp{GLSM} \cite{Witten:1993yc}, whose \ac{UV} description can accommodate a variety\footnote{Pun intended.} of \ac{IR} phases, including phase transitions between \acp{NLSM} and Landau-Ginzburg \acp{CFT} that would appear singular at the field-theoretic level. However, due to the nature of the geometry we seek to describe, the standard $(2,2)$ worldsheet supersymmetry is excessive. As in \cite{Anastasi:2026cus, Altavista:2026edv}, we shall consider $(1,1)$ worldsheet models consisting of real-scalar superfields. As a disclaimer, we emphasize that the class of models we will consider does not take into account the two different types of connected resolutions discussed in \cite{Baykara:2026gem, Altavista:2026evd}, which are useful to concoct rules that reproduce the various orientifolds of non-supersymmetric strings from the M-theoretic vantage point. \\

\noindent \paragraph{The local model.} The first model we consider is similar to those discussed in \cite{Altavista:2026edv}, albeit for a different purpose. The (internal) \ac{UV} worldsheet theory consists of three real-scalar superfields $X,Y,Z$ and the superpotential
\begin{eqaed}\label{eq:XYZ_model}
    W^\text{local}_0 = XYZ \, ,
\end{eqaed}
which leads to a classical moduli space with several branches. For our purposes we focus on the ``$Z$-branch'', where the \ac{VEV} $z \equiv \langle Z \rangle$ is varied. For $z \neq 0$ the two superfields $X, Y$ are massive, and localize at the origin $x=y=0$ where $x \equiv \langle X\rangle$ and $y \equiv \langle Y \rangle$. Thus, the \ac{IR} physics is that of a single real-scalar superfield. In contrast, when $z=0$ the two superfields $X,Y$ naively localize onto the singular algebraic locus defined by the equation
\begin{eqaed}\label{eq:xy_0}
    xy=0 \, ,
\end{eqaed}
and the \ac{IR} physics still describes string propagating in a single, albeit singular, internal dimension\footnote{Preserving the $Z$-branch is useful to describe the junction in a simpler fashion. Alternatively, one can include superpotential terms in \cref{eq:XYZ_model} that lift the $Z$-branch, as discussed e.g. in \cite{Altavista:2026edv}. Including such terms would modify the \ac{IR} structure without affecting the qualitative conclusions, as we will discuss later on in this section.}. This result highlights our motivation to consider \cref{eq:XYZ_model}: the locus described by \cref{eq:xy_0} is a local model for a one-dimensional wedge singularity, such as the one of $S^1 \vee S^1$; more precisely, it is a local model for a normal crossing, as opposed to two geometric circles tangent to each other. We shall not consider this possibility in this work. This local model can be deformed by replacing \cref{eq:XYZ_model} with
\begin{eqaed}\label{eq:XYZ_epsilon_model}
    W^\text{local}_\epsilon = (XY-\epsilon)Z \, ,
\end{eqaed}
whose scalar potential now contains the term $(xy-\epsilon)^2$. For $\epsilon \neq 0$, the superfield $Z$ is always massive and thus the \ac{IR} physics describes a single (smooth) internal dimension with
\begin{eqaed}\label{eq:xy_epsilon}
    xy=\epsilon \, .
\end{eqaed}
Then, from the \ac{UV} vantage point, changes in the deformation parameter along the worldsheet are encoded as interfaces. As in \cite{Altavista:2026edv}, $\alpha'$-corrections to this picture (which are quantum effects from the worldsheet perspective) can be systematically included, even exactly in some models. Specifically, as discussed in \cite{Hellerman:2004qa,Hellerman:2006ff,Anastasi:2026cus,Altavista:2026edv}, such (perturbative) corrections affect the kinetic terms but not the superpotential, thus preserving the geometric description we seek to describe in the \ac{IR}. Of course, in the singular case the latter will not comprise a \ac{NLSM}; rather, our strategy is to use the above procedures as means to \emph{define} wedge singularities in the worldsheet. \\

\begin{figure}[!ht]
    \centering
    \includegraphics[width=0.32\linewidth]{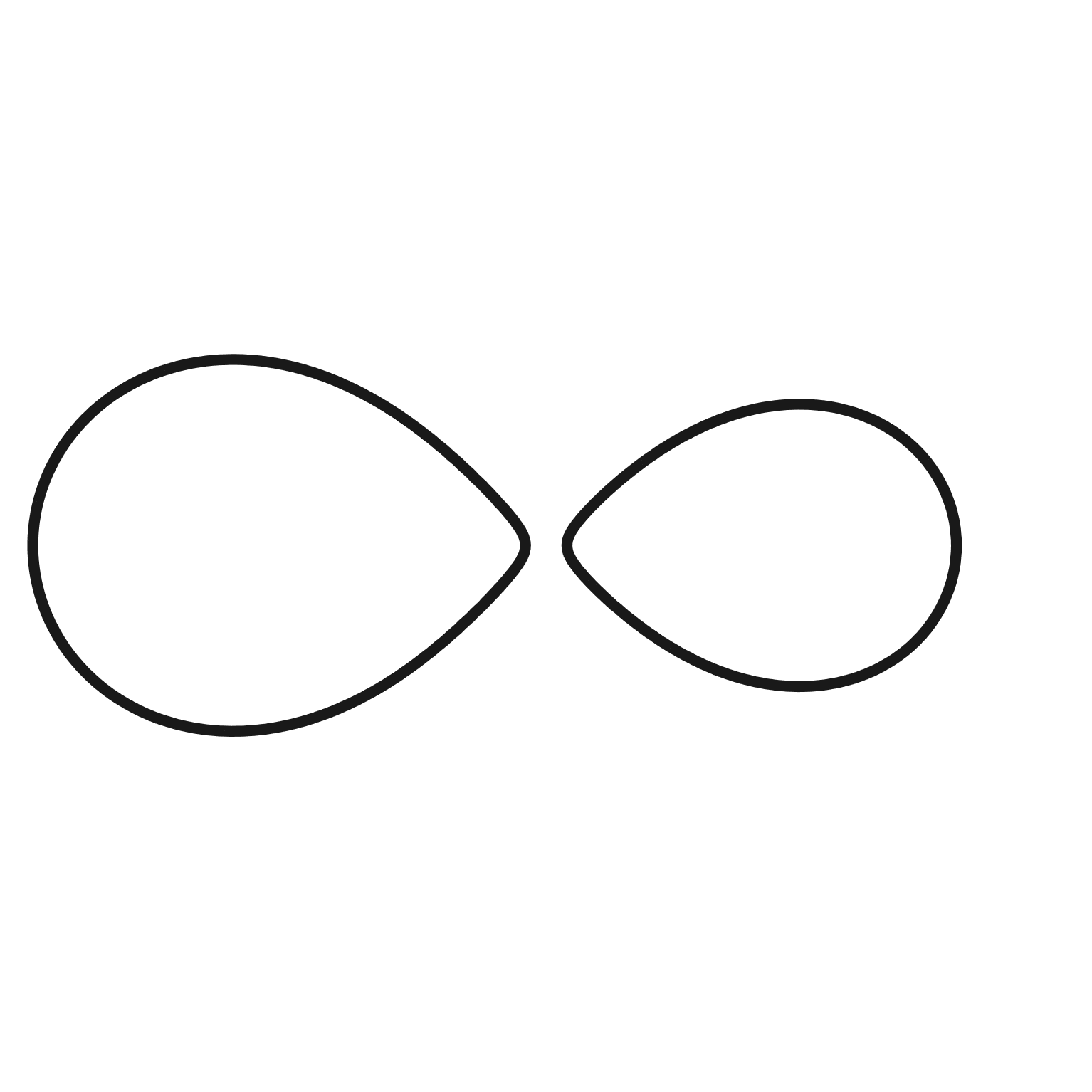}
    \includegraphics[width=0.32\linewidth]{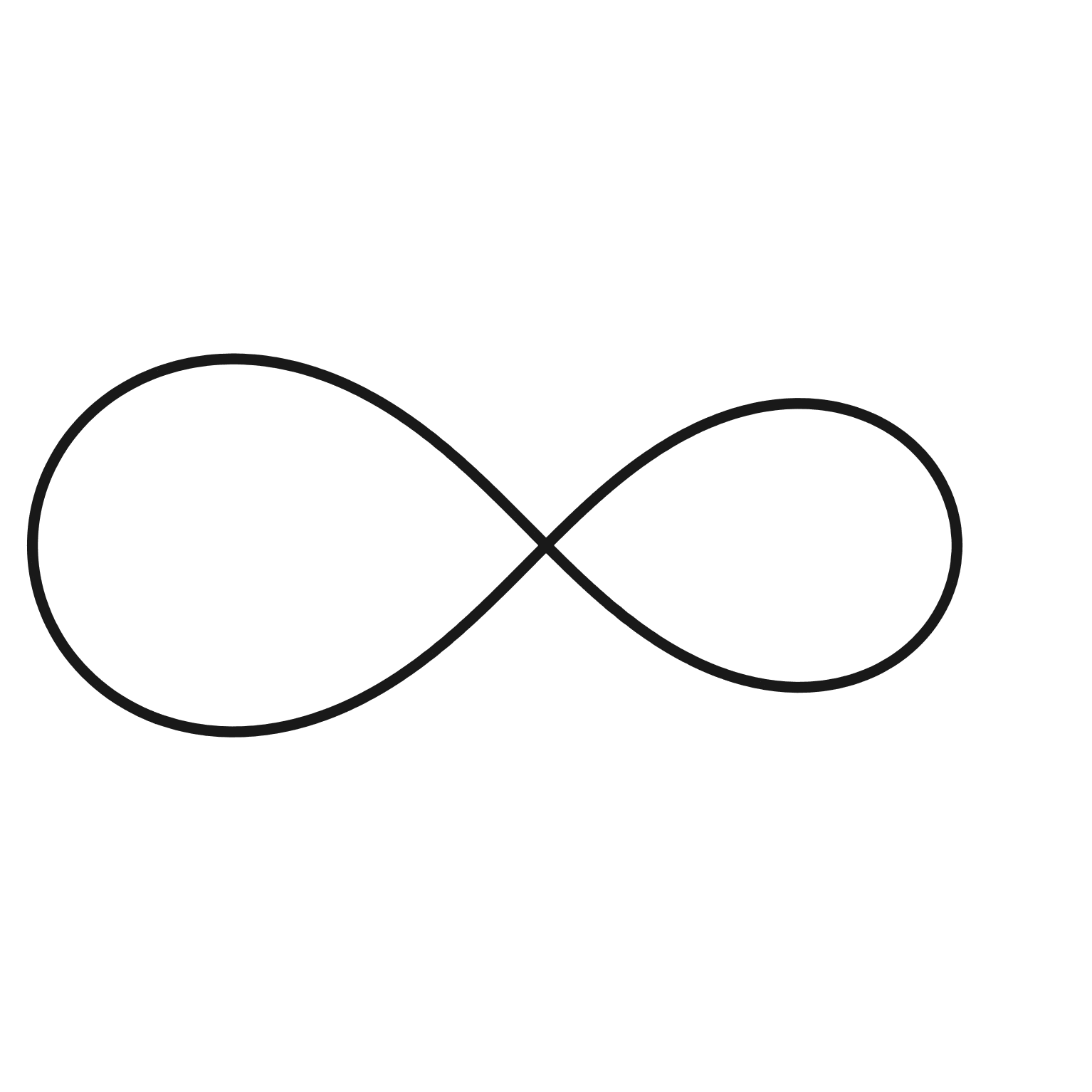}
    \includegraphics[width=0.32\linewidth]{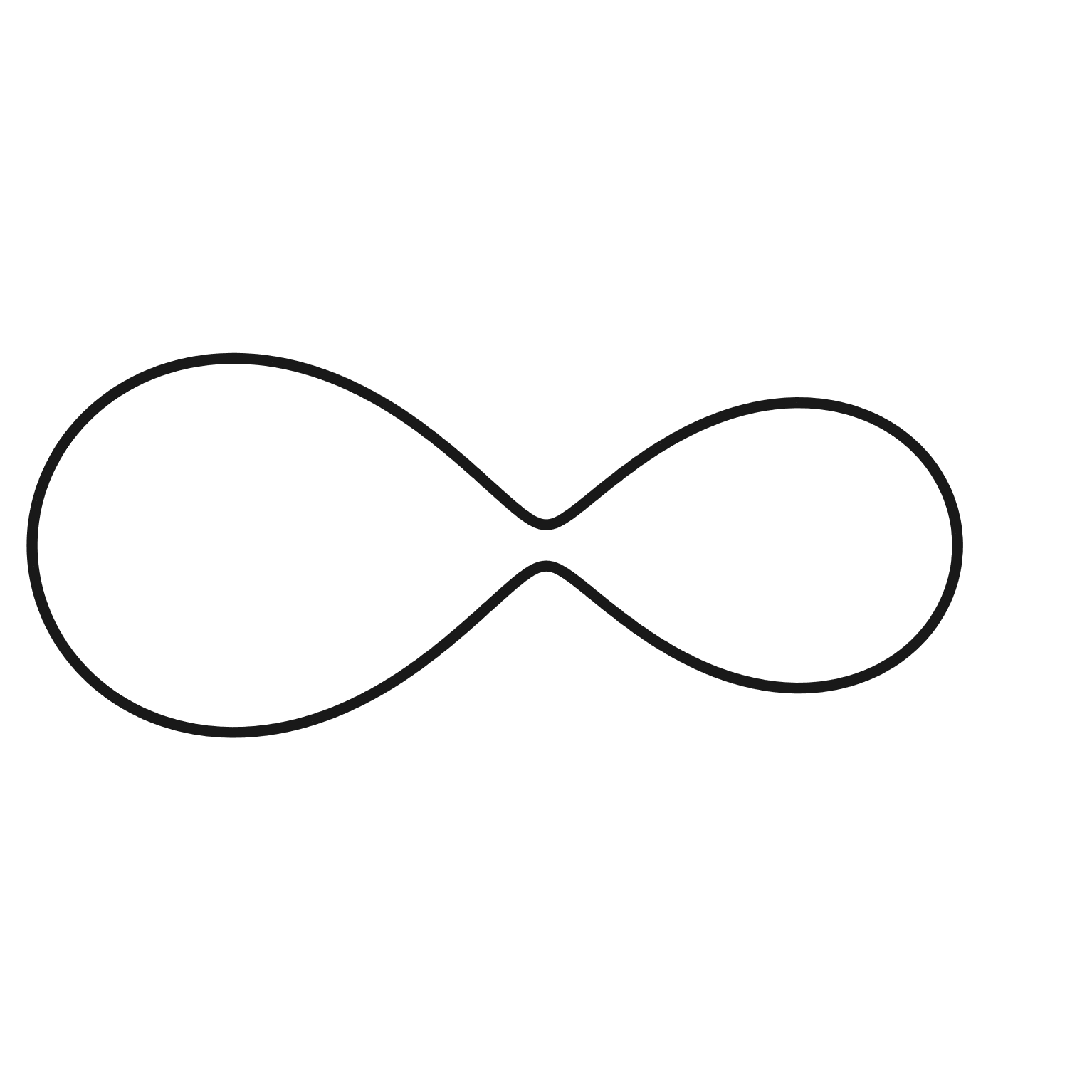}
    \caption{A depiction of the lemniscate (up to a coordinate rotation), generalizing \cref{eq:lemniscate} to the case of different radii $R_+ = 1.25R_-$, and its resolutions encoded in \cref{eq:W_lemniscate_different_radii}. From left to right: the disconnected resolution, the degeneration, and the connected resolution.}
    \label{fig:2d_lemniscate}
\end{figure}

\noindent \paragraph{The global model.} We now turn to a global description of the $S^1 \vee S^1$ singularity, starting with the case where the two components are equally large, say of characteristic radius $R$. To this end, a particularly convenient parametrization with a normal crossing\footnote{Although we expect that the qualitative conclusions be unaffected by these subtleties, in this work we focus on normal crossings due to their symmetric local behavior, relative to, e.g., tangency junctions.} (i.e. a lemniscate shape) is given by the algebraic equation
\begin{eqaed}\label{eq:lemniscate}
    x y = \frac{(x^2-y^2)^2}{R^2} \, ,
\end{eqaed}
which affords a deformation along the lines of \cref{eq:xy_epsilon} by subtracting $\epsilon$ from the left-hand side of \cref{eq:lemniscate}. The main advantage of this parametrization is that the complications due to compactness are isolated on the right-hand side, while the local structure of the singularity is isolated on the left-hand side. This feature will likely turn out to be instrumental in generalizing to other wedge singularities, as we discuss in more detail in \cref{sec:more_general_wedge_singularities}. The corresponding superpotential then takes the form
\begin{eqaed}\label{eq:W_lemniscate}
    W^\text{global}_\epsilon = \left(XY-\epsilon-\frac{(X^2+Y^2)^2}{R^2}\right)Z \, ,
\end{eqaed}
once more localizing the \ac{IR} dynamics onto the resolved lemniscate for $\epsilon \neq 0$. Note that, depending on the sign of $\epsilon$, the singularity is resolved according to the \ac{CRP} or \ac{DRP} \cite{Baykara:2026gem}, although the former has no orientation variant. This construction can be easily generalized to different radii $R_\pm$ for the two circles: up to a coordinate rotation, one can replace \cref{eq:W_lemniscate} with
\begin{eqaed}\label{eq:W_lemniscate_different_radii}
    W^\text{global}_\epsilon = \left((X^2+(Y-R_+)^2-R_+^2)(X^2+(Y+R_-)^2-R_-^2)+4R_+R_- (X^2 - \epsilon)\right)Z \, .
\end{eqaed}
See \cref{fig:2d_lemniscate} for a visual representation. Since this is used to obtain the correct light field content for some quotients \cite{Altavista:2026evd}, it would be interesting to investigate this aspect further. For the time being, we shall focus on the degenerate case of \cref{eq:W_lemniscate} with $\epsilon=0$, along with its simpler local realization of \cref{eq:XYZ_model}. In passing, we remark that this structure seemingly cannot be reproduced from a $(2,2)$ \ac{GLSM}, due to the lack of holomorphy. Similarly, in the $(1,1)$ case it is not clear to us whether continuous gaugings leading to D-term potentials of the desired form are possible or easily attained. \\

\subsection{Stringy corrections to the IR structure}\label{sec:quantum_corrections_IR}

As we have discussed above, for $\epsilon \neq 0$ the \ac{IR} physics is that of a $(1,1)$ \ac{NLSM} on $S^1$ or $S^1 \sqcup S^1$, depending on the sign of $\epsilon$. For $\epsilon=0$, the $Z$-branch of the classical moduli space receives significant $\alpha'$-corrections, as in \cite{Altavista:2026edv} (see also the references therein). This is related to the fact that $Z$ becomes massless at the junction point $x=y=0$ in \cref{eq:lemniscate}, as one can observe from \cref{eq:W_lemniscate} (for $\epsilon=0$). In this case, along the $Z$-branch the two other superfields $X, Y$ have masses proportional to $\abs{z}$, and thus the global structure away from the junction point---which is controlled by higher-order terms---is irrelevant. Hence, one can effectively focus on the local model of \cref{eq:XYZ_model}. The one-loop analysis carried out in \cite{Altavista:2026edv} then shows that the point $Z=0$ is renormalized at infinite distance in string frame. Furthermore, integrating out $X$ and $Y$ along the $Z$-branch generates a dilaton gradient, to the effect that the junction hosts a region of strong $\alpha'$-effects. Perhaps somewhat reminiscently of the well-known conifold transition, where wrapped D3-branes become massless, due to \cref{eq:string_coupling_map} non-perturbative degrees of freedom (invisible from the worldsheet) should also become important in this setting; a natural candidate would be D0-branes probing the junction, as we will discuss in \cref{sec:emergence}. At any rate, the presence of a strong-coupling effects, as well as $\alpha'$-effects, is consistent with the picture we outlined above: the connection between type IIA strings on $S^1 \vee S^1$ and type 0A strings on $S^1$, if any, should involve strongly coupled physics, and the wedge singularity cannot be described in the standard language of manifolds and (spacetime) fields. \\

\noindent \paragraph{Integrating out the massive sector.} In order to ascertain the above phenomena, we proceed at one-loop level in worldsheet perturbation theory. The one-loop effective action for $Z$ is obtained by expanding the superfields $X, Y$ to quadratic order around the relevant background, which in this case is $x=y=0$. The scalar potential $V = (\partial_X W)^2+(\partial_Y W)^2+(\partial_Z W)^2|_{X=x,Y=y,Z=z}$ provides the mass terms $z^2 x^2$ and $z^2 y^2$, so that integrating out the bosonic sector (formally) leads to
\begin{eqaed}\label{eq:tr_log_bosonic}
    S_\text{eff}^\text{bosonic} = -\frac{1}{2} \Tr \log (- \Box + z^2) = -\frac{1}{2} \int_{\Lambda_\text{UV}^{-2}}^\infty \frac{dt}{t} \, K_{-\Box+z^2}(t) \, ,
\end{eqaed}
where $\Lambda_\text{UV}$ is the \ac{UV} cutoff\footnote{Of course, we are ultimately interested in the \ac{IR} superconformal physics describing the actual string background(s) we seek. As such, the \ac{UV} cutoff ought to play no relevant role.}. We evaluate the relevant terms in \cref{eq:tr_log_bosonic} via the (diagonal, local) heat-kernel expansion of $K_{-\Box+z^2}(t) \equiv \text{tr} \, e^{-t(-\Box+z^2)}$ in terms of the Seeley-deWitt coefficients $\{a_{2k}\}$. For the complete operator $-\Box+z^2$, the expansion takes the form
\begin{eqaed}\label{eq:heat_kernel_expansion}
    K_{-\Box+z^2}(t) \overset{t \to 0^+}{\sim} \frac{\text{2}}{(4\pi t)^{\frac{d}{2}}} \int d^d\sigma \left(1 + t \, a_2[R_\text{ws}, z^2] + t^2 \, a_4[R_\text{ws}, z^2] + \dots\right)
\end{eqaed}
with $d=2$ the dimension of the worldsheet (parametrized by local coordinates $\sigma$); the overall factor of two arises by integrating out two fields. However, the corrections we seek involve a partial resummation of this expansion. The correction to the dilaton coupling is proportional to the worldsheet Ricci scalar $R_\text{ws}$, and thus includes the Seeley-deWitt coefficient $a_2[R_\text{ws}, z^2] = z^2 + R_\text{ws}/6$. More precisely, it arises resumming the contributions proportional to $R_\text{ws}$, which can be done by taking $z^2$ constant and factoring out the term $e^{-z^2 t}$ term from the heat kernel trace, leaving the Seeley-deWitt coefficient $a_2[R_\text{ws},0]=R_\text{ws}/6$. In two dimensions, this contribution to the effective action is accompanied by a logarithmic \ac{UV}-sensitivity, leading to a dilaton coupling of the form
\begin{eqaed}\label{eq:dilaton_correction}
    \phi(z) = \frac{2}{6} \int_{\Lambda_\text{UV}^{-2}}^\infty \frac{dt}{t} \, e^{-z^2 t} = \frac{2}{6} \log \frac{z}{\Lambda_\text{UV}} \, .
\end{eqaed}
To study the kinetic term, it suffices to work in a flat worldsheet, $R_\text{ws}=0$. Then the Seeley-deWitt coefficient $a_4$ is a total derivative, and the kinetic term is corrected by the Seeley-deWitt coefficient $a_6$, since it contains two worldsheet derivatives of the ``endomorphism'' $z^2$ \cite{Vassilevich:2003xt}, which up to boundary terms rearrange according to $a_6 \supset -(\partial z^2)^2/12$. This term is \ac{UV}-finite and suppressed by the background mass of the two superfields we are integrating out. Hence, as in \cite{Altavista:2026edv}, it is convenient to further introduce a background-fluctuation split $z(\sigma) = z_0 + \delta z(\sigma)$ and treat the constant term $z_0^2$ exactly in the heat-kernel expansion, leaving the dependence on worldsheet coordinates $\sigma$ into the fluctuation\footnote{Note that the ``fluctuation'' $\delta z$ is still considered to be part of the background, and is not integrated over.} $\delta z(\sigma)$. The upshot is that \cref{eq:tr_log_bosonic} contains an explicit factor of $e^{-z_0^2t}$. Since the final effective action must only depend on the combined superfield $Z$, one can read off the one-loop correction to the (bosonic) kinetic term $(\partial z)^2$ from the corresponding contribution
\begin{eqaed}\label{eq:kinetic_term_correction_schwinger}
    \frac{1}{2}\times\frac{2}{12} \int_{0}^{\infty} \frac{dt}{t^2} \, t^3 \, e^{-z_0^2t} \, (\partial z^2)^2 = \frac{1}{12} \, \frac{(\partial z^2)}{z_0^4} \overset{\delta z \ll z_0}{\sim} \frac{2}{6} \, \frac{(\partial z)^2}{z_0^2} \, .
\end{eqaed}
Since the full correction must only depend on the total field $z$, the above expression must resum into $(\partial z)^2/z^2$ as identified in \cite{Hellerman:2004qa,Hellerman:2006ff,Anastasi:2026cus,Altavista:2026edv}. To see this, we observe that on dimensional grounds, and on a flat worldsheet, the local part of the full heat-kernel trace can only contain two-derivative terms of the form
\begin{eqaed}\label{eq:resummed_heat_kernel}
    K_{-\Box+z^2}^\text{local}(t) = t^2 \, f(t z^2) \, (\partial z^2)^2
\end{eqaed}
for some function $f$, up to integration by parts in the effective action. Hence, the Schwinger integral of \cref{eq:tr_log_bosonic} produces the resummed kinetic structure $(\partial z)^2/z^2$, and the prefactor must match \cref{eq:kinetic_term_correction_schwinger} upon expanding around $z_0$.
\\

\noindent As for the fermionic sector, the heat-kernel computation is essentially analogous, with the proviso that the fluctuation operator be squared and brought into a canonical Laplace-like form. The contributions to the scalar potential, weighted by the Seeley-deWitt coefficient $a_0$, cancel against the bosonic ones, leaving a similar contribution to the renormalization of the kinetic term and dilaton gradient due to the different structure of the fluctuation operator, which indeed squares to the combination $-\Box - R_\text{ws}/4$. Hence, the relevant Seeley-deWitt coefficient $a_2$ is now given by
\begin{eqaed}\label{eq:fermionic_a2}
    a_2^\text{fermionic}[R_\text{ws},0] = \frac{R_\text{ws}}{6} - \frac{R_\text{ws}}{4} = - \, \frac{R_\text{ws}}{12} \, ,
\end{eqaed}
and the fermionic one-loop contribution $S_\text{eff}^\text{fermionic}$ brings along an overall minus sign with respect to \cref{eq:tr_log_bosonic}. Therefore, the total prefactor to \cref{eq:dilaton_correction} is $c/6$, where $c$ is the \ac{UV} central charge of the sector which is integrated out. Taking into account the prefactor of $c/6$ in the corrected target-space metric \cite{Hellerman:2004qa,Hellerman:2006ff,Anastasi:2026cus,Altavista:2026edv} (here we set $\alpha'=1$ for convenience), the canonical coordinate scales as $z_\text{can} = \sqrt{c/6} \, \log z$, resulting in a linear dilaton \ac{CFT} with slope $Q = \sqrt{c/6}$. Hence, since here $c=c_{X,Y}=3$, the central charge
\begin{eqaed}\label{eq:linear_dilaton_central_charge}
    c_\text{dilaton} = c_Z + 6Q^2 = \frac{3}{2}+3 = c_\text{UV}
\end{eqaed}
coincides with its \ac{UV} value. This makes intuitive sense, since the sector that was integrated out becomes massless at the junction. Hence, the \ac{IR} branch corresponding to the emergent degree of freedom at the junction is not critical; we will discuss how this problem can be resolved in \cref{sec:emergence}. Analogously to \cref{eq:kinetic_term_correction_schwinger}, the kinetic term for $z$ is also corrected, now from the structure of the Seeley-deWitt coefficients $a_4$ and $a_6$, since the squared Dirac operator now contains the combination $z^2+\partial\!\!\!/ z$. The $a_6$ structure produces the opposite contribution as in \cref{eq:kinetic_term_correction_schwinger}, while the $a_4$ coefficient contains $(\partial\!\!\!/ z)^2/2 = (\partial z)^2/z^2$. All in all, applying the argument around \cref{eq:resummed_heat_kernel}, this leads to the overall correction
\begin{eqaed}\label{eq:kinetic_correction_schwinger_fermions}
    \left(-\frac{1}{6}+\frac{1}{2} \times \frac{1}{2}\right) \frac{(\partial z)^2}{z^2} = \frac{1}{12} \, \frac{(\partial z)^2}{z^2}
\end{eqaed}
from the fermionic sector, matching the diagrammatic computation of \cite{Hellerman:2004qa,Hellerman:2006ff,Anastasi:2026cus,Altavista:2026edv}. \\

\noindent \paragraph{Localization of the emergent throat at the junction point.} Another way to see that the emergent throat is effectively localized at the junction point is to instead integrate out the $Z$ superfield in the local model of \cref{eq:XYZ_model} along the branches of the classical moduli space where $r^2 \equiv x^2+y^2 > 0$, i.e. away from the junction point. In this regime $Z$ has a mass proportional to $r$, and the structure of the Seeley-deWitt coefficient $a_2$ produces once more a term in the one-loop effective action that is analogous to \cref{eq:dilaton_correction}, namely
\begin{eqaed}\label{eq:dilaton_correction_xy}
    \phi(x,y) = \frac{c_Z}{6} \log \frac{r}{\Lambda_\text{UV}} \, .
\end{eqaed}
This shows that the region where strong $\alpha'$-effects affect the central charge in the \ac{IR} is indeed localized at the junction point. Similarly, using the global model of \cref{eq:W_lemniscate} for $\epsilon=0$ leads to the same conclusion, since the effective mass of the $Z$ superfield only vanishes at the junction point. As for the (bosonic) kinetic terms of the $X$ and $Y$ superfields, following an analogous logic as above one sees that they take the form $(\partial r \cdot \partial r)/r^2$, which (near the junction) reassembles into
\begin{eqaed}\label{eq:kinetic_corrected_xy}
  (\partial x)^2+(\partial y)^2 + \frac{\text{const.}}{x^2+y^2} \left(x^2 (\partial x)^2 + y^2 (\partial y)^2 + 2 x y (\partial x)\cdot (\partial y)\right) .
\end{eqaed}
Up to an overall positive prefactor, the components of the resulting $2 \times 2$ metric have eigenvalues 1 and $1 + \frac{\text{const.}}{r^2}$, showing the infinite-distance property of the junction. This property cannot be modified by the global geometry, and thus the effective (string-frame) length of the ``stringy'' geometry probed by the worldsheet is infinite. This property will play an important role in \cref{sec:stringy_corrections_light_species}, where the singularity will be resolved non-perturbatively. \\

\noindent \paragraph{Lifting the $Z$-branch.} As anticipated in a previous footnote, one might consider lifting the $Z$-branch by \emph{e.g.} adding a superpotential term proportional to $Z^3$, as in \cite{Altavista:2026edv}. For the degenerate model with $\epsilon=0$, this would keep $Z$ stabilized; integrating it out\footnote{Strictly speaking, naively there is a region in the \ac{UV} field space where the bosonic component $z$ is stabilized at a non-zero value. However, this region is obstructed in the \ac{IR}, as one can see by minimizing the resulting effective potential with respect to $x$ and $y$.}, the remaining superfields $X, Y$ would be subject to a scalar potential proportional to $x^2y^2$, while the fermionic components are massless. Flowing further toward the \ac{IR}, the two-branch structure of the classical moduli space is affected in a similar fashion as in the preceding paragraphs; namely, integrating out, say, $Y$ on the $X$-branch, the junction point is again pushed to infinite distance. As in the preceding cases, the \ac{IR} fermionic content matches the bosonic content in each branch as well. \\

\noindent \paragraph{Effective string coupling.} To conclude this section, we derive the effective string coupling in the \ac{IR}. Since the string coupling is exponential in the dilaton, the prefactors and scales that we have neglected in the above considerations can become important. Indeed, the superpotential in \cref{eq:W_lemniscate_different_radii} comes with a dimensionful coupling, while the induced dilaton in \cref{eq:dilaton_correction,eq:dilaton_correction_xy} comes with a \ac{UV} scale, both of which ought to be fixed. Moreover, both the superfield $Z$ and the radial mode in the $(X,Y)$ system must be integrated out to flow to the correct \ac{IR} (at least in the resolved models with $\epsilon \neq 0$), and they turn out to have the same effective mass, as we shall see. In order to address these issues, we begin by applying the framework we outlined above to a simple compactification of (say) type IIA strings on $S^1_R$. This can be achieved replacing \cref{eq:W_lemniscate_different_radii} with
\begin{eqaed}\label{eq:W_circle}
    W_\text{circle} = \lambda \, M_s^3 \, \, (X^2+Y^2-R^2) \, Z \, ,
\end{eqaed}
where $X,Y,Z$ are taken to have dimension of length and the coupling $\lambda$ has mass dimension one\footnote{This follows from the fact that the dimensionless superspace integral $\int d^2\sigma \, d\theta_+ d\theta_- \, W$ contains Grassmann factors whose mass dimension is the opposite of that of the Grassmann coordinates $\theta_\pm$ appearing in the superfield expansion in components.}. The normalization by the string scale $M_s^3$ arises from rescaling the canonical dimensionless scalar components by $M_s$ to obtain fields of mass dimension $-1$. The resulting scalar potential shows that the \ac{IR} physics localizes on the circle, and both the superfield $Z$ and the radial mode have a mass proportional to $\lambda \, M_s R$. Hence, each contributes to the effective dilaton as in \cref{eq:dilaton_correction,eq:dilaton_correction_xy}. Thus, the \ac{UV} dilaton $\phi_\text{UV}$ is shifted to its \ac{IR} value $\phi_\text{IR}$ according to
\begin{eqaed}\label{eq:IR_dilaton_circle}
    \phi_\text{IR} = \phi_\text{UV} + \frac{2}{4} \, \log\frac{\lambda \, M_s \, R}{\Lambda_\text{UV}} \, ,
\end{eqaed}
and the resulting nine-dimensional and ten-dimensional string couplings, related by $e^{2\phi_\text{IR}}=(g_s^\text{9d})^2=(g_s^\text{10d})^2M_s R$, are given in terms of the \ac{UV} worldsheet parameters by
\begin{eqaed}\label{eq:UV_IR_string_coupling}
    (g_s^\text{10d})^2 = e^{2\phi_\text{UV}} \, \frac{\lambda}{\Lambda_\text{UV}} \, .
\end{eqaed}
This fixes the corresponding couplings and scales (up to $O(1)$ numerical factors) for the case of our model of $S^1 \vee S^1$ compactification. In this case the superpotential in \cref{eq:W_lemniscate_different_radii} should be normalized by a prefactor $\lambda / (M_s^2R^2)$, where $R_\pm = R(1\pm T)$ and we take $T \ll 1$. This follows from the fact that ``tachyon condensation'' to $R_+=0$ (or $R_-=0$) leads back to the same \ac{IR} as a circle compactification \cite{Baykara:2026gem}, and the additional factor of $X^2+Y^2$ localizes to the constant $R^2$ in the \ac{IR}. Letting $F_\epsilon(X,Y)$ denote the function multiplying $Z$ in \cref{eq:W_lemniscate_different_radii}, we can observe that the effective masses of the superfield $Z$ and the radial mode are proportional to $\lambda M_s/R^2 \abs{\nabla F_\epsilon}$. For the superfield $Z$, this simply follows from the term $\lambda^2\norm{\nabla F}^2 z^2$ in the scalar potential. As for the radial mode, we consider the term $\lambda^2 F_\epsilon(x,y)^2$ in the scalar potential. Expanding $(x,y)$ around the \ac{IR} curve, for which $F_\epsilon=0$, the radial mode $\rho$ is identified by a perturbation of the form $(\delta x, \delta y) = \rho \nabla F_\epsilon/\norm{\nabla F_\epsilon}$, which is canonically normalized (up to factors of $M_s$). Then, one has
\begin{eqaed}\label{eq:radial_mass}
    F_\epsilon(x+\delta x,y+\delta y) \overset{M_s \rho \ll 1}{\sim} \rho \, \nabla F_\epsilon \cdot \frac{\nabla F_\epsilon}{\norm{\nabla F_\epsilon}} = \rho \, \norm{\nabla F_\epsilon} \, ,
\end{eqaed}
leading to the same effective mass of the superfield $Z$, and analogously for the fermionic superpartner of $\rho$. Hence, the local ten-dimensional string coupling $g_s^\text{local}$ is given by
\begin{eqaed}\label{eq:10d_effective_string_coupling}
    (g_s^\text{local})^2 = e^{2\phi_\text{UV}} \frac{\lambda}{\Lambda_\text{UV}} \, \frac{M_s \norm{\nabla F_\epsilon}}{R^2} = (g_s^\text{10d})^2 \, \frac{M_s \norm{\nabla F_\epsilon}}{R^2} \, .
\end{eqaed}
In addition, the \ac{IR} metric is corrected according to \cref{eq:kinetic_corrected_xy}, which again, upon restriction to the \ac{IR} curve, leads to an infinite-distance throat in string frame. \\

\noindent \paragraph{Summary of stringy corrections to the \ac{IR}.} All in all, following the above approach, we conclude that the effect of the wedge singularity is to produce a localized massless degree of freedom which develops a throat at infinite string-frame spacetime distance. However, due to the lack of criticality induced by the linear-dilaton gradient, it seems more appropriate to describe the junction in terms of the $Z$-branch, which is democratic with respect to the geometry in the $(x,y)$-plane, or in terms of the radial coordinate $r=\sqrt{x^2+y^2}$ upon integrating out $Z$. In particular, the latter normalizes canonically to a flat coordinate near the junction, and the dilaton gradient becomes linear as in the $Z$-branch. This approach democratizes the lack of criticality with respect to the singular geometry we strive to describe, but does not solve the issue in itself. In \cref{sec:emergence} we propose a resolution to this puzzle, which turns out to match some qualitative features of the weakly coupled type 0A string. Still, recalling that the putative duality between the type IIA and type 0A frames requires strong coupling effects in the former frame, we now investigate whether the spacetime description of this worldsheet phenomenon could be reproduced by integrating out D0-branes probing the wedge singularity, in the spirit of (some realizations of the) emergence proposal.

\section{Emergent type 0A tachyon from D0-branes}\label{sec:emergence}

Having discussed some dynamical aspects of our tentative worldsheet description, we now attempt to address strong-coupling effects. As mentioned in the preceding sections, these effects are necessary to describe weakly coupled type 0A strings from the type IIA frame, in view of \cref{eq:duality_M-IIA-0b,eq:string_coupling_map}. Once again, due to the lack of quantitative computational control, we stress that our ensuing considerations will remain at the level of parametric dependence, and will inevitably involve some speculative steps. The main goal of this section is to reproduce the tachyonic behavior of weakly coupled type 0A string theory from the type IIA side. To this end, we consider the potential produced by the lightest species in the limit $g_s^\text{IIA} \to +\infty$, subject to \cref{eq:string_coupling_map}. In \cref{sec:stringy_corrections_light_species} we will also consider instanton contributions, arguing that their structure reflects the strongly coupled nature of the limit we are considering. \\

\noindent \paragraph{Light species in the type IIA frame: D0-branes and wound branes.} In the type IIA frame, the strong-coupling limit leads to light D0-branes, whose mass (gap) $m_\text{D0} = M_s^\text{IIA}/g_s^\text{IIA} = 1/R_{11}$ is the \ac{KK} scale associated to the decompactification of the standard M-theory circle $S^1_{R_{11}}$ in \cref{eq:duality_M-IIA-0b}. However, in light of \cref{eq:string_couplings,eq:string_coupling_map}, we also find that $M_{11}R = (M_{11}R_{11})^{-2}$, which means that the size of the singular geometry $S^1 \vee S^1$ is small in this limit\footnote{One might be concerned about the fact that $R$ becomes sub-Planckian. However, as we shall discuss, the singularity remains parametrically smaller. This seems consistent with the overall idea of how this kind of ``quantum geometry'' ought to behave \cite{Baykara:2026gem, Altavista:2026evd}.}. Hence, in addition to D0-branes, fundamental type IIA strings wound around $S^1 \vee S^1$ also lead to light modes. In fact, since $M_{11}R_{11} \gg 1$, one might be tempted to interpret these modes as additional \ac{KK} species; however, $M_s^\text{0A}R_{11} = O(1)$ due to \cref{eq:string_coupling_map}, and thus the dual type 0A frame remains nine-dimensional. In addition, according to the M-theoretic picture of \cref{eq:duality_M-IIA-0b}, M2-branes wrapping $S^1 \vee S^1$ reduce to type 0A fundamental strings \cite{Baykara:2026gem}. Indeed, the corresponding mass scale is given by
\begin{eqaed}\label{eq:M2-brane_wrapped}
    \sqrt{M_{11}^3R} = M_s^\text{0A} \, ,
\end{eqaed}
and the accompanying oscillatory modes provide the emergent type 0A fundamental string in the dual frame. This picture is also consistent with \cref{eq:string_coupling_map}, which follows from the equi-dimensional condition $m_\text{D0} = M_s^\text{0A}$. These mass scales also match that of type IIA winding modes, which reads
\begin{eqaed}\label{eq:winding_match}
    (M_s^\text{IIA})^2R = M_s^\text{0A} \left(\frac{M_s^\text{IIA}}{M_s^\text{0A}}\right)^{\!2} \! (M_s^\text{0A}R) = M_s^\text{0A} \left(\frac{g_s^\text{IIA}}{g_s^\text{0A}}\right)^{\!\frac{2}{3}} \! g_s^\text{0A} = M_s^\text{0A} \, .
\end{eqaed}
We are thus led to investigate the contributions to the potential due to D0-branes and wound M2-branes in the type IIA frame.

\subsection{Emergence of tree-level tachyon from D0-branes}\label{sec:emergence_D0_potential}

\begin{figure}[!ht]
    \centering
    \includegraphics[width=0.8\linewidth]{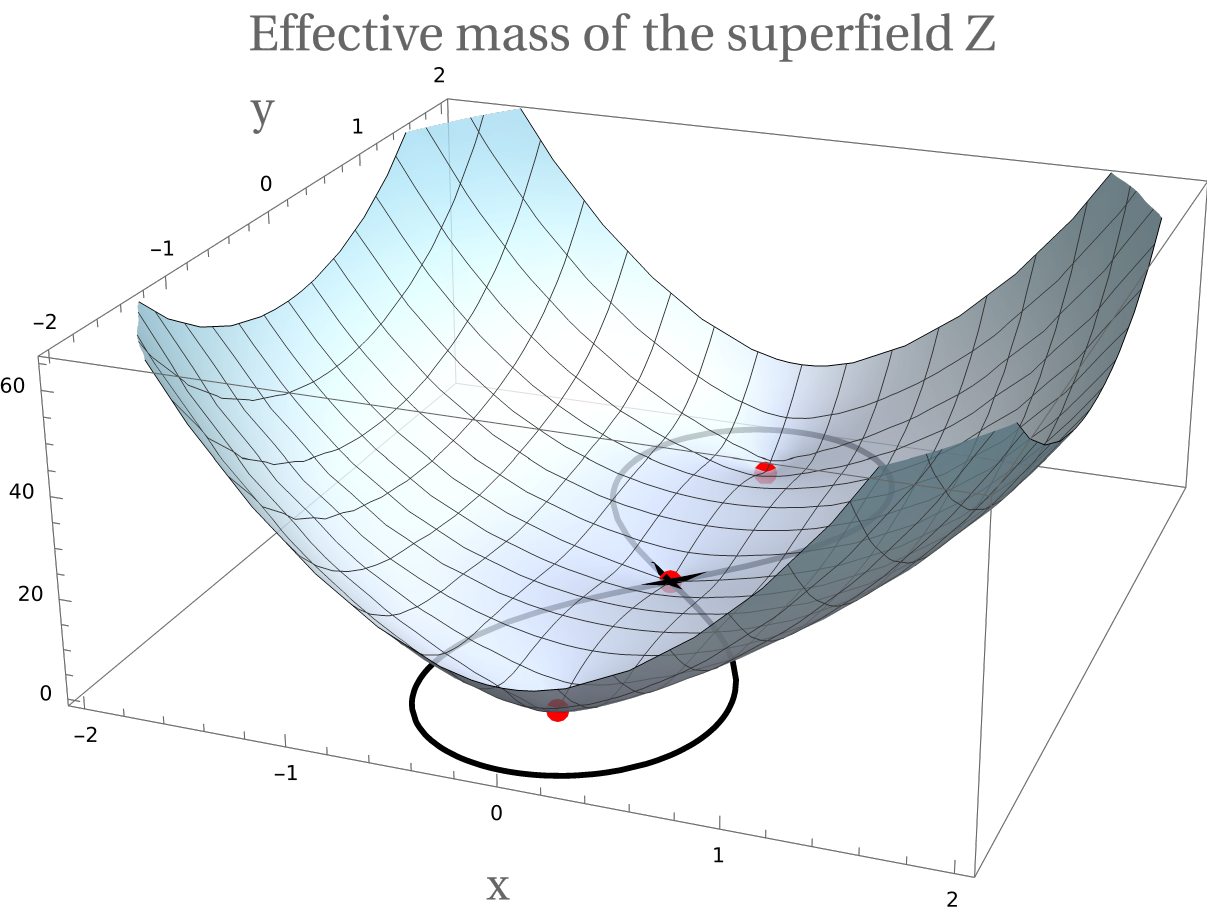}
    \caption{A plot of the effective mass in \cref{eq:Z_eff_mass} in string units, with $R_+ = R_-=1$. The global minima, where the effective mass vanishes, are shown as red dots; only the origin lies on the lemniscate, which in this case is described by \cref{eq:lemniscate} up to a coordinate rotation.}
    \label{fig:Z_eff_mass_plot}
\end{figure}

We begin by considering D0-branes. Since they are non-perturbative from the perspective of the type IIA frame, one is naturally led to conjecture that their contribution to the vacuum energy, or potential, reproduce tree-level physics in the putative dual type 0A frame, in the spirit of the (most recent incarnations of the) emergence proposal \cite{Heidenreich:2017sim,Heidenreich:2018kpg,Grimm:2018ohb,Corvilain:2018lgw,Blumenhagen:2023yws,Blumenhagen:2023tev,Blumenhagen:2023xmk,Hattab:2023moj,Blumenhagen:2024ydy,Blumenhagen:2024lmo,Hattab:2024thi,Hattab:2024ewk,Hattab:2024chf,Hattab:2024yol,Hattab:2024ssg,Artime:2025egu,Blumenhagen:2025zgf,Hattab:2025aok,Hattab:2026lho,Artime:2026dmm,Artime:2026kfq,Blumenhagen:2026rgu}. In particular, tree-level effects seems to be encoded---albeit somewhat mysteriously---by a renormalization scheme for one-loop Schwinger integrals involving a combination of minimal subtraction and zeta-function regularization \cite{Blumenhagen:2023yws,Blumenhagen:2023tev,Blumenhagen:2023xmk,Blumenhagen:2024ydy,Blumenhagen:2024lmo}. Thus, following this prescription, we compute the potential induced by integrating out D0-branes probing $S^1 \vee S^1$ using the heat-kernel method of \cref{eq:tr_log_bosonic,eq:heat_kernel_expansion}, according to
\begin{eqaed}\label{eq:V_D0_def}
    V_\text{D0} \equiv - \sum_{n \in \mathbb{Z}} \int^\infty_0 \frac{dt}{t^{1+\frac{d}{2}}} \, e^{-\pi n^2 m_\text{D0}^2 t} \bigg|_\text{prescription} \, ,
\end{eqaed}
where now $d=9$ in the case of interest, although the procedure works equally well in any dimension. Up to irrelevant positive numerical prefactors, we obtain\footnote{The overall minus sign in \cref{eq:V_D0_result} stems from the cancellation between the signs arising from the minimal subtraction and from $\zeta(-9) = -1/132$.}
\begin{eqaed}\label{eq:V_D0_result}
    V_\text{D0} \propto - \, \frac{M_{11}^9}{(g_s^\text{IIA})^6} \left(\frac{M_s^\text{IIA}\norm{\nabla F_\epsilon}}{R^2}\right)^{\!\!-9} ,
\end{eqaed}
where the local string coupling $g_s^\text{local}$ depends on the location of the D0-branes probing $S^1 \vee S^1$, due to the dilaton gradient of \cref{eq:10d_effective_string_coupling} induced by worldsheet effects. In type IIA units, and in the notation of \cref{eq:10d_effective_string_coupling}, the effective mass of the superfields that are integrated out is proportional to
\begin{eqaed}\label{eq:Z_eff_mass}
    \frac{M_s^\text{IIA}}{R^2} \, \sqrt{(\partial_x F_\epsilon)^2 + (\partial_y F_\epsilon)^2} \overset{r \to 0^+}{\sim} 4 M_s^\text{IIA} \, \frac{R_+ R_-}{R^2} \, r = 4(1-T^2) M_s^\text{IIA}r \, ,
\end{eqaed}
independently of the deformation parameter $\epsilon$. The asymptotic behavior in \cref{eq:Z_eff_mass} is the local behavior close to the junction, but it is also the global minimum of the effective mass\footnote{More precisely, there are other global minima in the $(x,y)$-plane. However, they do not lie on the geometry to which the superfields $X$ and $Y$ are constrained in the \ac{IR}. For a non-zero $\epsilon$, this can be shown explicitly, since the \ac{IR} phase is a free $(1,1)$ \ac{NLSM} on either $S^1_{R_+ + R_-}$ or $S^1_{R_+} \sqcup S^1_{R_-}$ depending on the sign of $\epsilon$. This is shown in \cref{fig:Z_eff_mass_plot}.}, as depicted in \cref{fig:Z_eff_mass_plot}. \\

\noindent Minimizing the potential with respect to the position of the D0-branes corresponds to minimizing the effective mass, which leads to a divergence for $\epsilon=0$. However, on physical grounds it is clear that this is not the correct procedure to define the potential. From the worldsheet perspective, the case $\epsilon=0$ leads to non-critical \ac{IR} physics due to the emergent throat. From the M-theoretic perspective, the quantum-geometric nature of the singularity rather suggests thinking of a Planckian deformation parameter $\abs{\epsilon} \lesssim M_{11}^{-2}$. This approach is consistent with both pictures: from the type 0A perspective, $(M_s^\text{0A})^2\epsilon = (g_s^\text{0A})^{\frac{2}{3}} M_{11}^2 \epsilon \ll 1$, and analogously in the type IIA frame when $g_s^\text{IIA} \ll 1$, while maintaining criticality in the \ac{IR} worldsheet \ac{CFT}. In other words, from the worldsheet side we are taking a limit in which the deformation parameter vanishes in the relevant string units as the string coupling is varied; this possibility is not available using solely \ac{CFT} ingredients, instead it requires specific string-theoretic input of spacetime interactions. From the M-theoretic perspective, a Planckian deformation of the singularity is precisely what is needed to account for the different resolution properties of type 0A fields \cite{Baykara:2026gem}. Hence, proceeding in this fashion, the potential is minimized at $r \approx \sqrt{\epsilon}$. Using \cref{eq:string_couplings,eq:string_coupling_map}, the resulting potential can be recast in the form
\begin{eqaed}\label{eq:V_D0_tree}
    V_\text{D0} \propto - \, \frac{M_{11}^9}{(M_{11}R_{11})^{\frac{27}{2}}(M_{11}^2\epsilon)^{\frac{9}{4}}(1-T^2)^{\frac{9}{2}}} = - \, \frac{(M_s^\text{0A})^{10}R_{11}}{(1-T^2)^{\frac{9}{2}}} \left(\frac{g_s^\text{0A}}{(M_{11}\sqrt{\epsilon})^3}\right)^{\!\frac{3}{2}} \, .
\end{eqaed}
In order to reproduce the tree-level potential in the putative weakly coupled type 0A dual frame, which comes from the compactification on $S^1_{R_{11}}$, we must take
\begin{eqaed}\label{eq:epsilon_matching}
    M_{11}\sqrt{\epsilon} = (g_s^\text{0A})^{\frac{7}{9}} \ll 1 \, .
\end{eqaed}
This also implies that $M_s^\text{IIA} \sqrt{\epsilon} \ll 1$ in the limit we are considering; in fact, the singularity befittingly looks small in all frames, and remains parametrically smaller than the whole space, since $\epsilon \ll R$. This effect is perturbatively invisible in the type IIA frame, as apparent from \cref{eq:epsilon_matching,eq:string_coupling_map}. Most importantly, the dependence on $T$ has no tadpole and is tachyonic, and has the (parametrically) correct tree-level mass $\abs{m_T} = M_s^\text{0A}$, once the canonical normalization by powers of $M_s^\text{0A}$ is taken into account. Moreover, the M-theoretic picture of \cite{Baykara:2026gem} is most relevant when $g_s^\text{0A} = O(1)$, whereby \cref{eq:epsilon_matching} entails a Planckian singularity, the natural expectation for a quantum geometry. Albeit very speculative, the fact that the rather simple scaling limit embodied by \cref{eq:epsilon_matching} exists and satisfies all these requirements at once is non-trivial and is consistent with the quantum-geometric picture of \cite{Baykara:2026gem,Altavista:2026evd,Baykara:2026vdc,Altavista:2026brr}. Even taking this matching for granted, at first glance there seems to be an issue: at $T=0$, weakly coupled type 0A strings predict a vanishing potential, and its value can be expected to be relevant due to the presence of dynamical gravity. However, a second glance provides a possible resolution for this puzzle within the equi-dimensional limit spelled out in \cref{eq:string_coupling_map}, as we will outline shortly in \cref{sec:stringy_corrections_light_species}. \\

\noindent \paragraph{Wound fundamental type IIA strings.} Let us recall that the light spectrum of D0-branes is accompanied by light winding modes of type IIA fundamental strings, which from our preceding considerations we might expect to dualize to the type 0A fundamental string oscillations. The contribution of these modes to the potential, due to the supersymmetry-breaking orbifold implied by the M-theoretic resolution property of the gravitino \cite{Baykara:2026gem}, is naively expected to scale according to $V_\text{winding} \propto ((M_s^\text{IIA})^2R)^9 = (M_s^\text{0A})^9$ from the schematic one-loop expression involving a sum over sectors of the Narain lattice of signature $(1,1)$,
\begin{eqaed}\label{eq:one-loop_winding}
    V_\text{winding}^\text{one-loop} & = -\frac{(M_s^\text{IIA})^9}{2}\sum_{\sigma \in \text{sectors}} \int_{\mathbb{H}^2/\text{PSL}(2,\mathbb{Z})} \frac{d\tau d\overline{\tau}}{(\text{Im} \, \tau)^2} \, Z^\text{IIA}_\sigma(\tau, \overline{\tau}) \, \Gamma_{\sigma}^{\text{Narain}(1,1)}(\tau, \overline{\tau}; R) \\
    & \overset{M_s^\text{IIA}R \ll 1}{\sim} \text{const.} \times (M_s^\text{IIA})^9 \, \int^\infty_{O(1)} \frac{dt}{t^{1+\frac{9}{2}}} \, e^{-\frac{\pi t}{M_s^\text{IIA}R}} \propto ((M_s^\text{IIA})^2R)^9 = (M_s^\text{0A})^9 \, .
\end{eqaed}
This indeed reproduces the one-loop potential in the type 0A frame. Moreover, the different ways that strings can wind around $S^1 \vee S^1$ can produce a dependence on $T$, leading to one-loop corrections to the tree-level mass of the type 0A tachyon, as well as the expected dilaton tadpole (in the Einstein frame). Similar considerations are expected from M2-branes wrapping $S^1 \vee S^1$ on dimensional grounds, due to \cref{eq:M2-brane_wrapped}. While these naive estimates reproduce the one-loop contribution in the putative type 0A dual frame, due to the lack of supersymmetric protection mechanisms there seems to be nothing preventing these contributions from canceling the offset in \cref{eq:V_D0_tree} and provide a complete match to the type 0A potential at tree and one-loop level. In fact, we now discuss some corrections to the above expressions which arise from the extrapolation of the worldsheet analysis of \cref{sec:strings_on_wedge_singularities}. While a quantitative match cannot be expected at strong coupling, these effects provide a proof of principle for the possibility of a complete non-perturbative duality, in the spirit of \cite{Baykara:2026gem,Altavista:2026evd,Baykara:2026vdc,Altavista:2026brr}.

\subsection{Stringy corrections to the winding potential}\label{sec:stringy_corrections_light_species}

The calculation leading to \cref{eq:V_D0_result,eq:V_D0_tree} crucially relies on the presence of a dilaton gradient induced by worldsheet effects. Similarly, the naive potential in \cref{eq:one-loop_winding} is expected to be modified by the kinetic terms induced by worldsheet effects, as discussed in \cref{eq:kinetic_corrected_xy}. Of course, at strong coupling there is no \emph{a priori} reason to trust the ensuing computations at the quantitative level. Still, in the spirit of \cite{Baykara:2026gem,Altavista:2026evd,Baykara:2026vdc,Altavista:2026brr}, they provide a proof of concept at the qualitative level that there are in fact mechanisms that produce significant corrections to \cref{eq:one-loop_winding}, possibly removing the tree-level offset of \cref{eq:V_D0_result}. Here we account for these corrections by computing the proper length of the curve $F_\epsilon=0$, in the notation of \cref{eq:radial_mass,eq:10d_effective_string_coupling}, with respect to the metric in \cref{eq:kinetic_corrected_xy}. We keep $\epsilon \neq 0$, anticipating a (logarithmic) divergence as $\epsilon \to 0$. Letting $(x(s),y(s))$ be a parametrization of the curve, we are led to compute the line element $\abs{dr/r}$ (with a prefactor of $1/M_s^\text{IIA}$ in our conventions). The tree-level tachyon mass is reproduced by \cref{eq:V_D0_result,eq:epsilon_matching}, and is unaffected by these corrections since D0-branes do not wrap the internal space; hence, we simplify matters by setting $T=0$, i.e. we consider $R_\pm = R$. Then, the symmetry of problem allows us to parametrize half of the curve by solving the equation $F_\epsilon=0$ with respect to $x$, and among the real branches $x = x_\pm(y)$ we pick the positive one for convenience. The residual symmetry of the geometry further allows us to restrict to $y > 0$. Then, we compute
\begin{eqaed}\label{eq:line_element}
    \abs{\frac{dr}{r}} = \frac{2y \, dy}{R^2+2y^2+\epsilon-R\sqrt{R^2+2y^2+\epsilon}} \, .
\end{eqaed}
The proper length of the curve is given by (four times) the integral of \cref{eq:line_element} within the relevant extrema of integration. More precisely, we should also include the tree-level term in the line element; however, the asymptotic behavior in the singular limit is unchanged. The complete integrand is simply obtained by summing the tree-level term (which is just unity) and the one-loop term in quadrature. While the integration extrema are calculable, their general expressions are rather complicated; they are also useless of our purposes, since we ultimately want to take the limit in which $\epsilon \ll R^2$. Then, the upper integration bound can be replaced by $y_\text{max} = 2R$, while the lower bound depends on the sign of $\epsilon$ and can be replaced by $y_\text{min} = R \, s_\text{min}(\epsilon/R^2)$, with
\begin{eqaed}\label{eq:y_min}
    s_\text{min}(\alpha) = \sqrt{\abs{\alpha}} \, \theta(-\alpha) = \begin{cases}
0 \, , \qquad \epsilon>0 \, , \\
\sqrt{-\alpha} \, , \qquad \epsilon<0
\end{cases}
\end{eqaed}
to leading order, where (here and in the following) $\theta$ denotes the Heaviside theta function. Putting things together, and including the prefactor of $1/4$ in the full one-loop metric, the proper length $L_\epsilon$ of the geometry as $\epsilon \to 0$ is given by
\begin{eqaed}\label{eq:proper_length}
    L_\epsilon \overset{\epsilon \ll R^2}{\sim} \frac{1}{M_s^\text{IIA}} \, \mathcal{J}\left(\frac{\epsilon}{R^2}\right) & \, , \\
    \mathcal{J}(\alpha) \equiv \int_{s_\text{min}(\alpha)}^2 \frac{2s \, ds}{1+2s^2 + \alpha - \sqrt{1+2s^2 + \alpha}} = & \int_{\abs{\alpha} \, \theta(-\alpha)}^4 \frac{\sqrt{1+2\zeta}+1}{\sqrt{1+2\zeta}} \frac{d\zeta}{2\zeta+\alpha}\, .
\end{eqaed}
In the final equality we have rationalized the denominator and performed the change of integration variables $\zeta = s^2$. As expected, the integral diverges logarithmically as $\alpha \to 0$. This can be ascertained by analyzing the local behavior of the integrand around zero, which gives the dominant contribution
\begin{eqaed}\label{eq:integral_asymptotics}
    \mathcal{J}(\alpha) \overset{\alpha \to 0}{\sim} \int_{\abs{\alpha} \, \theta(-\alpha)}^{O(1)} \frac{2 d\zeta}{2\zeta + \alpha} = \log \frac{1}{2\abs{\alpha} \, \theta(-\alpha) + \alpha} + O(1) = \log \frac{1}{\abs{\alpha}} + O(1) \, .
\end{eqaed}
This result was expected on the grounds that \cref{eq:line_element} yields a logarithmic divergence at the radial origin, and the minimal distance is proportional to $\sqrt{\epsilon}$, as in \cref{eq:V_D0_tree}. As anticipated above, while these computations are not necessarily reliable at the quantitative level in the regime that ought to match with a dual type 0A frame, they nonetheless provide a proof of principle that a mechanism correcting \cref{eq:one-loop_winding} is at play, leaving open the possibility that \cref{eq:V_D0_tree,eq:one-loop_winding} (along with other contributions) combine to yield the correct tree-level (and possibly one-loop) physics of weakly coupled 0A strings. \\

\noindent \paragraph{Stringy corrections to D0-instantons.} By the same token as above, the Euclidean action of D0-instantons wrapping the internal space also receives corrections in the type IIA frame. These amount to multiplying \cref{eq:line_element} by $M_s^\text{IIA}/g_s^\text{local}$ before integration, where the local string coupling is given by \cref{eq:10d_effective_string_coupling}. Using \cref{eq:proper_length} and including the factor of $1/4$ from the metric, the resulting instanton action takes the form
\begin{eqaed}\label{eq:D0-instanton_action}
    & S_\text{D0-instanton} = \frac{2^{-\frac{7}{4}}}{g_s^\text{IIA}\sqrt{M_s^\text{IIA}R}} \, \mathcal{I}\left(\frac{\epsilon}{R^2}\right) = \left(2 \,g_s^\text{0A}\right)^\frac{1}{4} \, \mathcal{I}\left(\frac{\epsilon}{R^2}\right) \, , \\
    \mathcal{I}(\alpha) & \equiv \int_{\abs{\alpha} \, \theta(-\alpha)}^4 \frac{d\zeta}{\left((1+\sqrt{\zeta})(\sqrt{1+2\zeta+\alpha}-1)\right)^\frac{1}{4}(1+2\zeta+\alpha-\sqrt{1+2\zeta+\alpha})} \\
    & = \int_{\abs{\alpha} \, \theta(-\alpha)}^4 \frac{\left(\sqrt{1+2\zeta+\alpha}+1\right)^\frac{5}{4}}{\left(1+\sqrt{\zeta}\right)^\frac{1}{4}\sqrt{1+2\zeta+\alpha}} \, \frac{d\zeta}{\left(2\zeta+\alpha\right)^\frac{5}{4}} \, .
\end{eqaed}
A local analysis along the lines of \cref{eq:integral_asymptotics} then yields
\begin{eqaed}\label{eq:instanton_asymptotics}
    \mathcal{I}(\alpha) \overset{\alpha \to 0}{\sim} 2^\frac{1}{4} \int_{\abs{\alpha} \, \theta(-\alpha)}^{O(1)} \frac{2 d\zeta}{\left(2\zeta + \alpha\right)^\frac{5}{4}} = 4\left(\frac{2}{\abs{\alpha}}\right)^\frac{1}{4} \,  + O(1) \, ,
\end{eqaed}
which is a stronger suppression of the exponential $e^{-S_\text{D0-instanton}}$ relative to \cref{eq:integral_asymptotics}. This was expected, based on \cref{eq:10d_effective_string_coupling}, but the complete D0-instanton action in \cref{eq:D0-instanton_action} also contains a factor of $(g_s^\text{0A})^\frac{1}{4}$. Combining the above results and using \cref{eq:epsilon_matching}, one finally arrives at
\begin{eqaed}\label{eq:D0-instanton_asymptotics}
    S_\text{D0-instanton} \overset{\epsilon \ll R^2}{\sim} \sqrt{2} \left(g_s^\text{0A}\right)^{\frac{7}{36}} \overset{g_s^\text{0A} \to 0^+}{\ll} 1 \, ,
\end{eqaed}
which indeed highlights the strongly coupled nature of the limit we are considering. Once again, albeit not necessarily reliable in the regime at stake, these calculations are at least a qualitative consistency check that this regime exhibits the required physical features, among which a lack of suppression for instanton contributions in the type IIA frame.

\section{Worldsheet cobordisms and wedge singularities}\label{sec:cobordisms_singularities}

Having discussed in detail the most interesting case of $S^1 \vee S^1$, in this section we lay down a generalized framework to study other kinds of wedge singularities from the worldsheet perspective. We first formulate a straightforward extension of the methods presented in \cref{sec:strings_on_wedge_singularities}. Then, we apply topological tools (specifically, topological modular forms and the Stolz-Teichner conjecture) to analyze wedge singularities in terms of worldsheet cobordisms.

\subsection{More general wedge singularities}\label{sec:more_general_wedge_singularities}

In principle, the construction presented in \cref{sec:strings_on_wedge_singularities} can be generalized to other kinds of wedge singularities, at least if they allow a real-algebraic description. To this end, let $\mathbf{X} = (X_i)_{i=1, \dots, n}$ denote a collection of $n$ real-scalar superfields, with lowest components $\mathbf{x} = (x_i)_{i=1, \dots, n}$, and let $F_\epsilon(\mathbf{x})=0$ describe the (resolved) geometry we wish to probe via a $(1,1)$ worldsheet theory. If $f_\epsilon(\mathbf{x})$ describes the local behavior of the singular locus, we can add the additional real-scalar superfield $Z$ and consider the superpotentials
\begin{eqaed}\label{eq:general_W}
    W_\epsilon^\text{local} & = f_\epsilon(\mathbf{X}) \, Z \, , \\
    W_\epsilon^\text{global} & = F_\epsilon(\mathbf{X}) \, Z \, .
\end{eqaed}
In the degenerate case $\epsilon=0$, for which the equation $f_0(\mathbf{x})=0$ defines a singular locus, we expect the \ac{IR} physics to produce a similar behavior as that studied in \cref{sec:strings_on_wedge_singularities}. However, given the greater freedom allowed by $(1,1)$ worldsheet supersymmetry, a more systematic study is needed to ascertain this in general. The simplest strategy to obtain wedge singularities is to construct $F_\epsilon$ by embedding the disjoint union $M \sqcup M$ of two copies of a (compact, closed) real-algebraic variety $M$ into its canonical ambient space, and let $\epsilon \to 0^+$ describe the limit in which they become tangent. However, this approach has two clear shortcomings: firstly, the wedge singularity is generically not a normal crossing; secondly, $\epsilon < 0$ does not describe a smooth connected manifold. In order to make progress at this greater level of generality, the (real) algebraic geometry of blowups and exceptional divisors may prove helpful; intuitively, the role played by the extra superfield $Z$ in \cref{sec:strings_on_wedge_singularities} may be carried out by superfields describing string propagation on exceptional divisors. \\

\begin{figure}[!ht]
    \centering
    \includegraphics[width=0.32\linewidth]{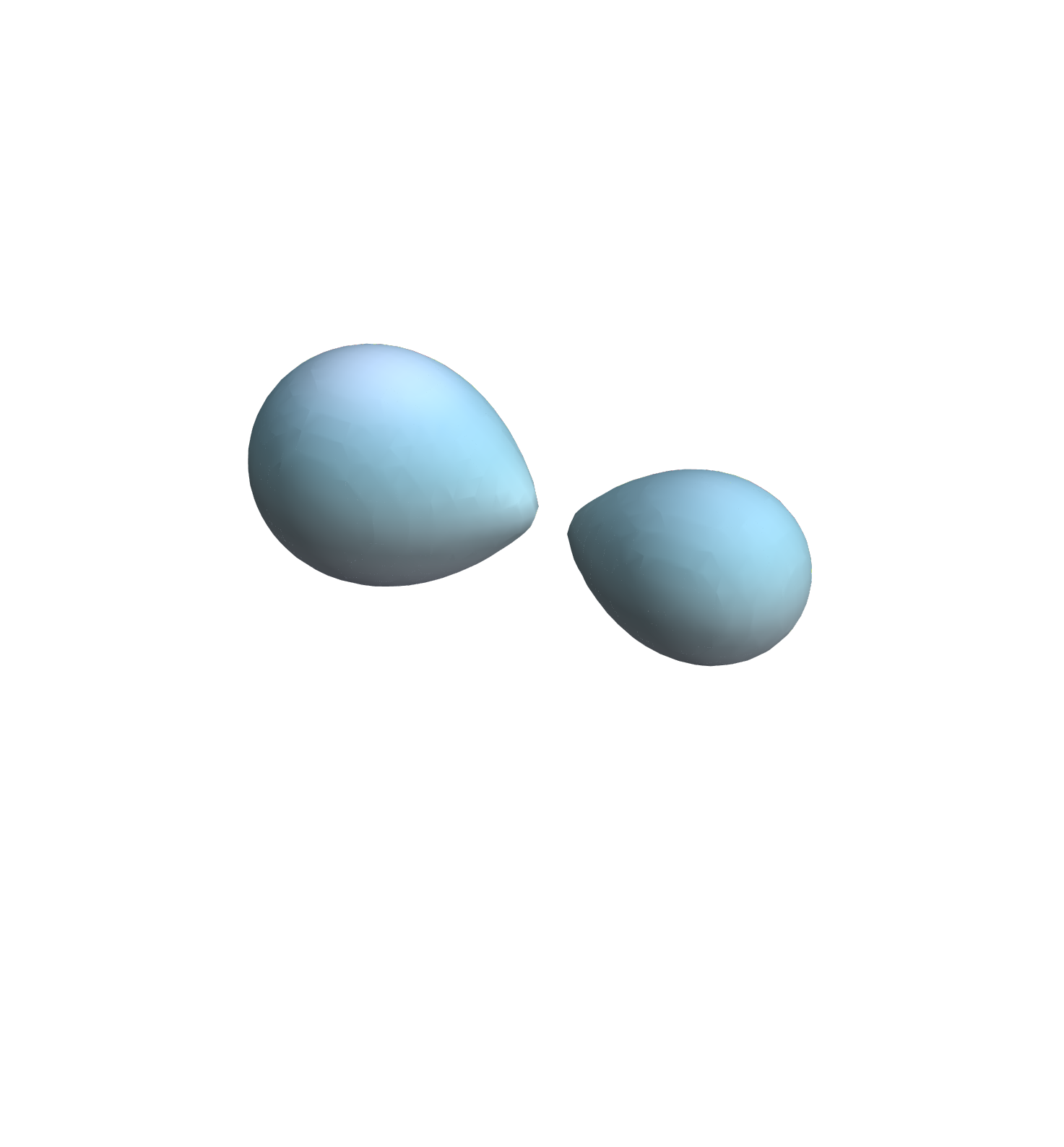}
    \includegraphics[width=0.32\linewidth]{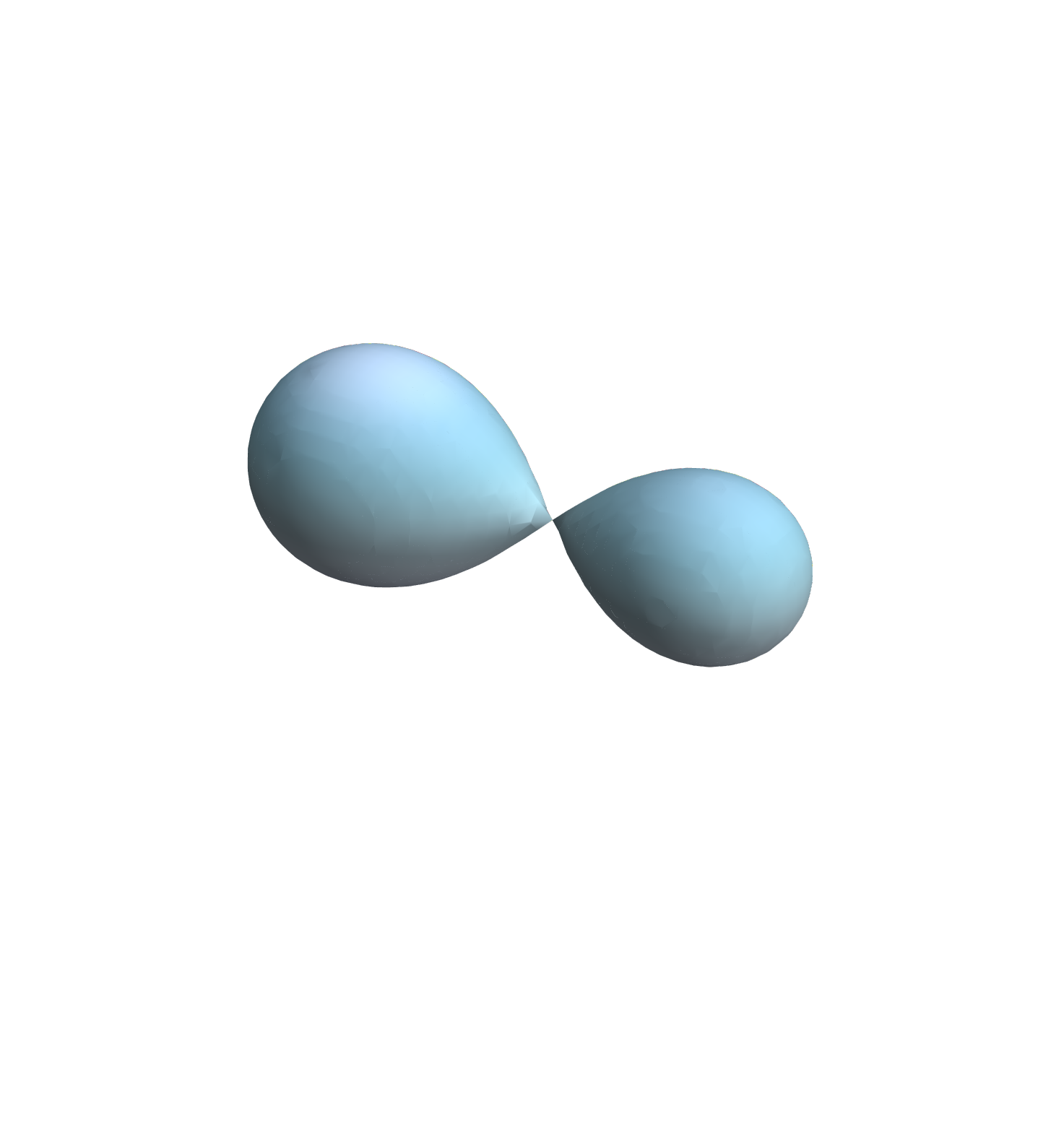}
    \includegraphics[width=0.32\linewidth]{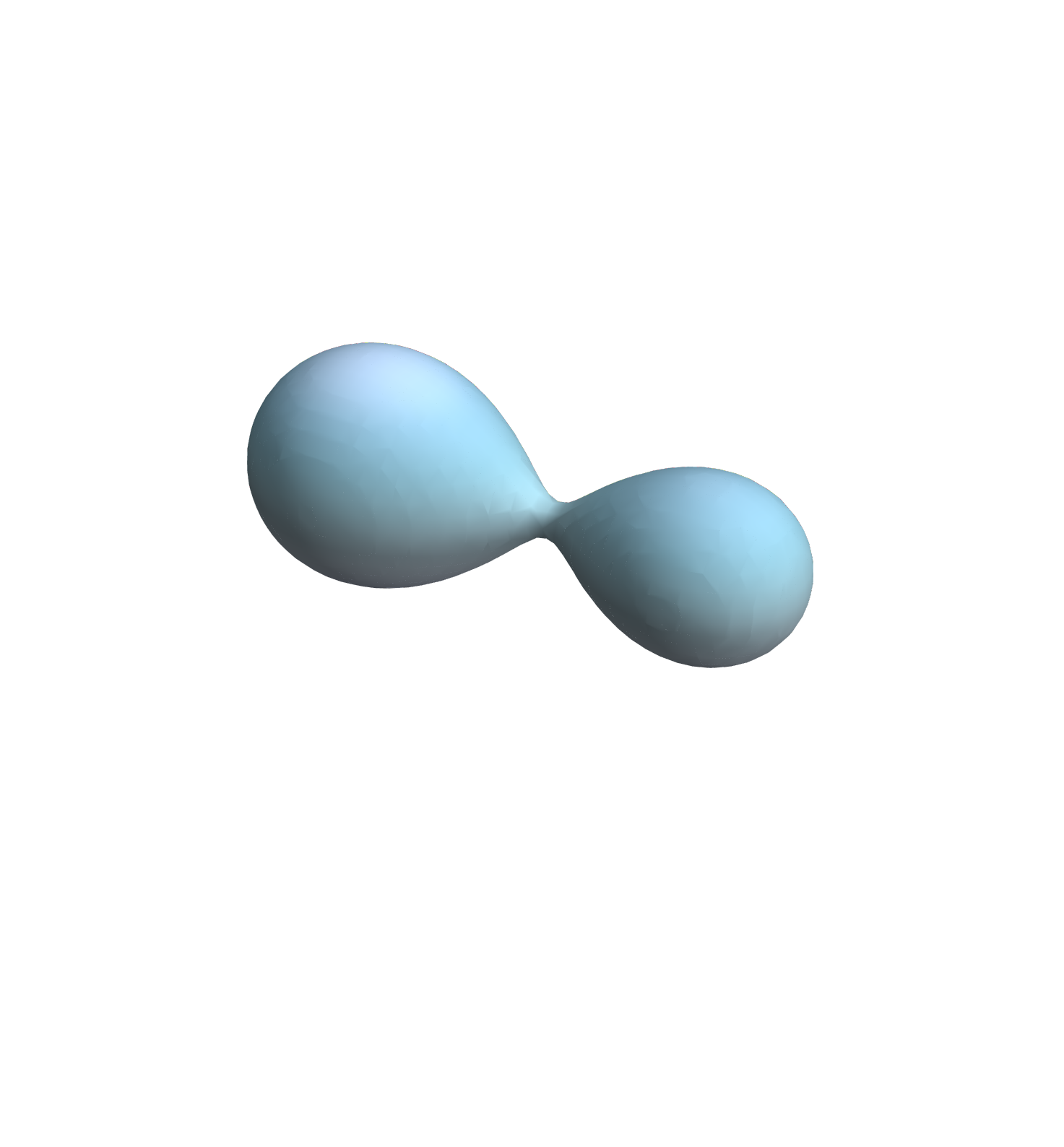}
    \caption{A depiction of the $d=2$ generalized lemniscate, deforming \cref{eq:tangent_spheres} to a normal crossing, with radii $R_+ = 1.25R_-$ and its resolutions. From left to right: the disconnected resolution, the degeneration, and the connected resolution.}
    \label{fig:3d_lemniscate}
\end{figure}

\noindent Alternatively, since real algebraic geometry is somewhat less rigid than complex algebraic geometry due to its lack of holomorphy, a more ``pedestrian'' strategy to solve the above issues may pay off. Namely, if $M$ is smooth, bringing to tangency two disjoint copies of $M$ embedded into their canonical ambient space, the local structure of the wedge singularity is that of two tangent (osculating) spheres of dimension $d \equiv \dim M$. Thus, locally, the deformation parameter $\epsilon$ controlling the connected and disconnected resolutions of $M \vee M$ can be introduced by embedding $S^d \vee S^d$ into $\mathbb{R}^{d+1}$ with a normal crossing at the junction point. This treatment should also apply to any pair $M,N$ of such manifolds joined into $M \vee N$. Letting $(\mathbf{x},y)$ denote canonical coordinates on $\mathbb{R}^{d+1}$, with $y$ the coordinate along which the embedded copies of $M$ are brought together, the local behavior of the tangent configuration is simply described by the equation $y^2=0$, up to higher-order homogeneous corrections. We wish to deform this local model into a double cone along $y$, described by the equation $y^2 = \norm{\mathbf{x}}^2$. For $d=1$, this geometry reduces to the one discussed in \cref{sec:strings_on_wedge_singularities} up to a coordinate rotation. In order to achieve a normal crossing for any $d$, generalizing \cref{eq:W_lemniscate_different_radii}, consider the equation
\begin{eqaed}\label{eq:tangent_spheres}
    (\norm{\mathbf{x}}^2 + (y-R_+)^2 - R_+^2)(\norm{\mathbf{x}}^2 + (y+R_-)^2 - R_-^2) = 0
\end{eqaed}
describing tangent spheres of radii $R_\pm$. Locally near the origin, the left-hand side is given by $-4R_+ R_- y^2$. Hence, adding $4R_+R_- \norm{x}^2$ to the left-hand side of \cref{eq:tangent_spheres} deforms the wedge singularity into a normal crossing. To wit, for $d=1$, after a $45^\circ$ rotation the resulting equation matches \cref{eq:lemniscate} for $R_+ = R_- \equiv R$ and generalizes it to $R_+ \neq R_-$ as well as $d > 1$. Finally, the deformation parameter $\epsilon$ can be introduced simply by subtracting $4R_+R_- \epsilon$ from the left-hand side (at least insofar as $\epsilon \ll R_\pm^2$). As an example, \cref{fig:3d_lemniscate} depicts the case $d=2$ of the generalized lemniscate which deforms \cref{eq:tangent_spheres} into a normal crossing. \\

\noindent Putting everything together, we thus arrive at a $(1,1)$ worldsheet model with real-scalar superfields $\mathbf{X},Y,Z$ and (a family of) superpotential(s) given by
\begin{eqaed}\label{eq:W_spheres}
    W_\epsilon = \left((\norm{\mathbf{X}}^2 + (Y-R_+)^2 - R_+^2)(\norm{\mathbf{X}}^2 + (Y+R_-)^2 - R_-^2) + 4R_+R_- (\norm{X}^2 \!- \epsilon) \right) Z
\end{eqaed}
describing $S^d \vee S^d$ and resolutions thereof. Following the same analysis as in \cref{sec:strings_on_wedge_singularities}, by axial symmetry it follows that the conclusions do not change; then, since the \ac{IR} physics of the degeneration at $\epsilon=0$ into a wedge singularity is localized at the junction, the case of $S^d \vee S^d$ should generalize to $M \vee N$ for any pair $M, N$ of smooth, compact, closed manifolds. In particular, the global superpotential in \cref{eq:general_W} should reduce to \cref{eq:W_spheres} with $R_\pm$ the radii of the osculating spheres to $M$ and $N$ at the junction point. If one so chooses, one can also multiply the term $4R_+R_- \norm{X}^2$ by (a factor proportional to) $\epsilon$, trading the normal crossing for the tangent singularity.

\subsection{Worldsheet cobordisms and topological modular forms}\label{sec:worldsheet_cobordisms}

The dynamical analysis of \cref{sec:strings_on_wedge_singularities} can be complemented by some kinematic considerations. In the spirit of \cite{Altavista:2026edv}, it might be fruitful to examine the family of deformed worldsheet theories describing wedge singularities, along with its resolutions, in terms of cobordisms of worldsheet theories. Focusing on equivalence classes under deformations---or simply ``deformation classes''---allows us to leverage tools from algebraic topology. Indeed, the space of two-dimensional supersymmetric \acp{QFT} is expected to arrange into a ($E_\infty$) ring spectrum \cite{Gukov:2018iiq, Tachikawa:2025flw, Tachikawa:2025awi}; specifically, for minimal supersymmetry, this spectrum is conjectured by Stolz and Teichner \cite{Stolz_Teichner_2004, Stolz:2011zj} to correspond to (a specific, periodic variant of) \ac{TMF}. This connection traces back to the overlap between robust properties of two-dimensional supersymmetric \acp{QFT} and elliptic cohomology \cite{Witten:1986bf, Witten:1987cg}, and the place that \ac{TMF} occupies as a ``universal'' elliptic cohomology theory, obtained as the global sections of the structure spectral sheaf over the spectral Deligne-Mumford stack of elliptic curves \cite{10.1007/978-3-0348-9078-6_49}. \\

\noindent \paragraph{The setup for worldsheet cobordisms across wedge singularities.} The well-supported Stolz-Teichner conjecture has been used to great effect in heterotic strings, where it finds its natural home \cite{Tachikawa:2021mby, Tachikawa:2021mvw}. For our purposes, in the absence of a suitable analog for $(1,1)$ theories (including the possibility of D-branes), we simply regard our worldsheet models of wedge singularities as $(0,1)$ theories. In each of their non-singular \ac{IR} phases, these theories are $(1,1)$ \acp{NLSM} on a target manifold $M$; we thus regard them as $(0,1)$ \acp{NLSM} with target space $M$ coupled to a Fermi multiplet valued in the tangent bundle $TM$. This is a special case of the general $(0,1)$ \acp{NLSM} with target space $M$ coupled to a Fermi multiplet valued in a vector bundle $V \to M$. In particular, depending on the \ac{IR} branch of moduli space, $M = M_1 \sqcup M_2$ or $M = M_1 \# M_2$ is the disjoint union or the connected sum of two compact, closed manifolds $M_1, M_2$. We will now investigate the possibility of connecting these branches via worldsheet cobordisms in the sense of \ac{TMF}. \\

\noindent \paragraph{Deformation classes of \ac{IR} phases.} In the standard case of $(0,1)$ \acp{NLSM} on a target space $M$ of dimension $\dim M=d$ equipped with a string structure\footnote{For our purposes, a string structure on a spin manifold $M$ is a trivialization of the canonical generator $\lambda = \frac{p_1}{2} \in H^4(B\text{Spin},\mathbb{Z})$ pulled back to $M$ via the classifying map $TM : M \to B\text{Spin}$.}, the string orientation of \ac{TMF}
\begin{eqaed}\label{eq:string_orientation_tmf}
    M\text{String} \to \text{TMF}
\end{eqaed}
produces maps
\begin{eqaed}\label{eq:cobordism_map}
    \Omega^\text{string}_d = \pi_d(M\text{String}) \to \pi_d(\text{TMF}) \equiv \text{TMF}_d \to \text{MF}[\Delta^{-1}]_{\frac{d}{2}} \, ,
\end{eqaed}
where $M\text{String}$ is the Thom spectrum\footnote{Strictly speaking, we should be talking about Madsen-Tillman spectra $MT\xi$ for tangential structures $\xi$. The difference will not matter in this work.} associated to the classifying space $B\text{String}$ of the 2-group $\text{String}$, defined as the homotopy fiber over the map $\lambda : B\text{Spin} \to K(\mathbb{Z},4)$ corresponding to the generator class $\lambda = \frac{p_1}{2} \in H^4(B\text{Spin}, \mathbb{Z})$. The last arrow in \cref{eq:cobordism_map} is the (integral) Witten genus $\phi_\text{W}$, whose image lies in (integral) modular forms with the discriminant $\Delta = \eta^{24}$ inverted\footnote{This inversion is absent if one considers the connective version of \ac{TMF}, denoted tmf. Here $\eta$ is the Dedekind eta function.} to allow for poles at the cusp. Letting a two-dimensional \ac{CFT} $T$ with $(0,1)$ supersymmetry and central charges $(c_\text{L}, c_\text{R})$ represent a deformation class $[T] \in \text{TMF}$, its gravitational anomaly $\nu \equiv 2(c_\text{R}-c_\text{L}) \in \mathbb{Z}$ is the \ac{TMF} degree of $[T]$, and
\begin{eqaed}\label{eq:witten_genus}
    \phi_\text{W}([T]) = \eta(q)^{\nu}\text{tr}_{\mathcal{H}_\text{Ramond}(T)} (-1)^{F_\text{R}} q^{L_0 - \frac{c_\text{L}}{24}}
\end{eqaed}
is given by a trace over the right-moving Ramond sector $\mathcal{H}_\text{Ramond}(T)$ of $T$. In \cref{eq:witten_genus}, the Dedekind factor $\eta(q)^\nu$ is needed to ensure that $\phi_\text{W}([T]) \in \text{MF}[\Delta^{-1}]_\frac{\nu}{2}$ is a modular form (with integral coefficients, and possibly poles at the cusp). For $T = \sigma(M)$ a \ac{NLSM} with target space $M$, the gravitational anomaly $\nu = \dim M \equiv d$ and \cite{Witten:1986bf, Witten:1987cg, Tachikawa:2021mby, Tachikawa:2021mvw}
\begin{eqaed}\label{eq:witten_genus_NLSM}
    \phi_W([\sigma(M)]) = \left(q^{-\frac{1}{24}}\eta(q)\right)^d \!\int_M \widehat{A}(TM) \wedge \text{ch} \left(\bigotimes_{n \geq 1} S_{q^n} TM\right) ,
\end{eqaed}
where $\widehat{A}$ is the Dirac genus and $S_x(TM) \equiv \bigoplus_{k \geq 0} x^k \, \text{Sym}^k(TM)$ is a (formally graded) direct sum of symmetric tensor powers of $TM$. Similarly, we will need its antisymmetric counterpart $\Lambda_x V \equiv \bigoplus_{k \geq 0} x^k \, \Lambda^k(V)$. When the \ac{NLSM} on a spin target manifold $M$ is coupled to a Fermi multiplet valued in a vector bundle $V \to M$ of (even) rank $r$ with $\lambda(TM) = \lambda(V)$, which we shall denote $T=\sigma(M,V)$, the appropriately modular generalization of \cref{eq:witten_genus_NLSM} is \cite{Schellekens:1986xh,Witten:1986bf, Witten:1987cg, Ando:2009av}
\begin{eqaed}\label{eq:witten_genus_NLSM_twisted}
    \phi_W([\sigma(M,V)]) = \left(q^{-\frac{1}{24}}\eta(q)\right)^{d-r} \!\int_M \widehat{A}(TM) \wedge \text{ch}\left(\Delta^{(-)}(V) \bigotimes_{n \geq 1} \left[S_{q^n} TM \otimes \Lambda_{-q^n} V \right]\right) ,
\end{eqaed}
where $\Delta^{(-)}(V) \equiv \Delta_-(V) \ominus \Delta_-(V)$ is the virtual difference of the positive and negative chirality spinor bundles $\Delta_\pm(V)$. Notably, \cref{eq:witten_genus_NLSM_twisted} can be thought of as a ``twisted'' $S^1$-equivariant index formula \cite{ando2001wittengenusequivariantelliptic, dai2026ellipticcherncharacterselliptic}. For $d=r$, the dependence on $q$ in \cref{eq:witten_genus_NLSM_twisted} disappears \cite{Witten:1986bf, Witten:1987cg} and $\phi_\text{W}([\sigma(M,V)]) = \chi(V)$ gives the Euler characteristic of $V$. This is particularly easy to see for $V = TM$, for which the twisted-string condition $\lambda(TM) = \lambda(V)$ is trivial, $\text{ch}\left(\Delta^{(-)}(TM)\right) = e(TM)/\widehat{A}(TM)$ and the $q$-dependent factors cancel out. Indeed, in terms of Chern-Weil curvature two-forms $F_V, F_{TM}$ the formal structure appearing in \cref{eq:witten_genus_NLSM_twisted} is
\begin{eqaed}\label{eq:formal_structure}
    \prod_{n\geq 1}\frac{1-q^n e^{iF_V}}{1-q^n e^{iF_{TM}}} \, ,
\end{eqaed}
showing the cancellation explicitly for $V=TM$. \\

\noindent \paragraph{Worldsheet cobordisms across wedge singularities.} The upshot of the above considerations is that, if $\phi_\text{W}([T_1]) \neq \phi_\text{W}([T_2])$, the theories $T_1$ and $T_2$ are not worldsheet-cobordant. In particular, we are interested in the case where $T_1 = \sigma(M_1 \sqcup M_2,T(M_1 \sqcup M_2))$ and $T_2 = (M_1 \# M_2, T(M_1 \# M_2))$, which ought to be the endpoints of a worldsheet cobordism across the singular space $M_1 \vee M_2$. Returning to the initial setting of $S^1 \vee S^1$, the integral Witten genus of the \acp{NLSM} corresponding to each resolution of the wedge singularity vanishes. However, since \cite{Gukov:2018iiq, Tachikawa:2021mby}
\begin{eqaed}\label{eq:TMF_groups}
    & \text{TMF}_{-1} = 0 \, , \\
    & \text{TMF}_0 = \mathbb{Z}[x] \, , \\
    & \text{TMF}_1 = \mathbb{Z}_2[x] \, ,
\end{eqaed}
where $x$ is mapped into the $j$-invariant by the last arrow in \cref{eq:cobordism_map}, there are additional invariants to consider \cite{bunke2009secondaryinvariantsstringbordism, Tachikawa:2023lwf, Tachikawa:2023nne, Tachikawa:2024ucm, Tachikawa:2025flw}. In this case, the mod-2 elliptic genus \cite{Tachikawa:2023nne} is relevant, since its value on $[\sigma(S^1)]$ generates $\mathbb{Z}_2 \subset \text{TMF}_1$. According to this characterization, the \ac{IR} branches of the worldsheet theory describing strings propagating on $S^1 \vee S^1$ are not worldsheet-cobordant, since $[\sigma(S^1 \# S^1 \simeq S^1)]$ has non-trivial mod-2 elliptic genus, while $[\sigma(S^1 \sqcup S^1)] = 2[\sigma(S^1)]$ has trivial mod-2 elliptic genus. However, this is only true at the level of the minimally supersymmetric \acp{NLSM}; we must take into account that the relevant \acp{NLSM} for our purposes have $(1,1)$ supersymmetry, and decompose into \acp{NLSM} with $(0,1)$ supersymmetry coupled to a Fermi multiplet valued in the tangent bundle of the target space. These \acp{QFT} have gravitational anomaly $\nu = 0$, and thus according to \cref{eq:TMF_groups} the relevant invariant is the integral Witten genus. As discussed below \cref{eq:witten_genus_NLSM_twisted}, we have that
\begin{eqaed}\label{eq:witten_genus_euler}
    \phi_\text{W}([\sigma(M,TM)]) = \chi(M)
\end{eqaed}
is simply the Euler characteristic of $M$. Then, since
\begin{eqaed}
    & \chi(M_1 \sqcup M_2) = \chi(M_1) + \chi(M_2) \, , \\
    & \chi(M_1 \# M_2) = \chi(M_1) + \chi(M_2) - (1+(-1)^d) \, ,
\end{eqaed}
where $d = \dim M_1 = \dim M_2$, we conclude that a worldsheet cobordism across a wedge singularity does not exist for $d$ even. For the most interesting case for us, namely $d=1$, the class
\begin{eqaed}\label{eq:product_theories}
    [\sigma(S^1,TS^1)] = [\sigma(S^1)] \times [\text{Fermi multiplet}] \in \text{TMF}_0
\end{eqaed}
is the product of free theories in degrees $1$ and $-1$. Since $\text{TMF}_{-1}=0$ \cite{Gukov:2018iiq}, as in \cref{eq:TMF_groups}, we learn that $[\sigma(S^1,TS^1)] = 0$ is the trivial degree-zero class, and thus there is no obstruction to a worldsheet cobordism from $S^1 \sqcup S^1$ to $S^1 \# S^1 \simeq S^1$ due to the additional Fermi multiplet completing the field content of a $(1,1)$ real-scalar superfield. It would be very interesting to assess whether an obstruction exist and be detectable by a more refined notion of worldsheet cobordism for $(1,1)$ supersymmetry. However, as we have pointed out in \cref{sec:strings_on_wedge_singularities}, there seems to be a dynamical obstruction due to the infinite-distance region localized at the junction. At the worldsheet level, this means that the description given in \cref{sec:strings_on_wedge_singularities} is still that of a ``non-compact worldsheet cobordism'', as in \cite{Anastasi:2026cus, Altavista:2026edv}. At the spacetime level, the presence of strong-coupling effects due to \cref{eq:string_coupling_map} indicates that, even if a worldsheet cobordism across $S^1 \vee S^1$ were realizable in a compact fashion, there would likely be additional, non-perturbative consistency conditions to account for, in order that a duality with weakly couple type 0A strings be feasible.

\section{Conclusions}\label{sec:conclusions}

In this paper we investigated a number of aspects of string theory on wedge singularities, especially in the context of the M-theoretic duality web proposed in \cite{Baykara:2026gem,Altavista:2026evd,Baykara:2026vdc,Altavista:2026brr}. Having introduced a scaling limit achieving a putative weakly coupled type 0A dual frame via a strong-coupling limit of type IIA strings on $S^1 \vee S^1$ in \cref{sec:dualities_setup}, we studied dynamical aspects in \cref{sec:strings_on_wedge_singularities,sec:emergence} and topological aspects in \cref{sec:worldsheet_cobordisms}, as part of the generalized framework of \cref{sec:more_general_wedge_singularities}. In particular, in \cref{sec:emergence} we speculated on the possibility that type IIA D0-branes yield the tree-level potential in the type 0A frame, and found that the emergence proposal---specifically as applied in \cite{Blumenhagen:2024ydy,Blumenhagen:2026rgu}---leads to the correct behavior of the type 0A tachyon if the wedge singularity is resolved non-perturbatively in a specific scaling limit. \\

\noindent The upshot of the dynamical analysis of \cref{sec:strings_on_wedge_singularities} is that, independently of which superfields are integrated out to investigate the \ac{IR} physics of the worldsheet, one finds an emergent throat localized at the junction, which ends up at infinite (string-frame) spacetime distance due to $\alpha'$-corrections. This feature turned out to be essential in reproducing the tachyonic behavior of the potential and providing a physical mechanism to remove the overall offset of the potential. \\

\noindent As for the topological considerations in \cref{sec:worldsheet_cobordisms}, although for the case of $S^1 \vee S^1$ the analysis seems to be inconclusive, in general our results show that the wedge singularity cannot always be ``crossed'' via traditional (compact) cobordisms, even when manifolds are abstracted to worldsheet theories. It is possible that a suitable analog of the Stolz-Teichner conjecture for $(1,1)$ theories could hold further insights; however, dynamically, the strongly coupled physics produced by the singularity in order to match to a weakly coupled type 0A frame still requires a non-perturbative treatment beyond mere worldsheet methods. As already mentioned, \cref{sec:emergence} we made the first steps in this direction, investigating the interplay between D0-branes and the type 0A frame from the perspective of the emergence proposal. \\

\noindent \paragraph{Outlook.} Let us conclude with some thoughts on future directions. From the worldsheet perspective, it would be interesting to further investigate other aspects of type 0A physics, such as the doubling of \ac{RR} sectors which we preliminarily discussed in \cref{sec:K-theory}. From the M-theoretic perspective, it would be interesting to develop novel, if partially qualitative, tools to investigate the strongly coupled physics hosted by such wedge singularities (such as non-commutative or derived/spectral algebraic geometry) in order to possibly collect additional evidence for the their consistency and use in the unveiling of the non-supersymmetric web of string dualities. For instance, a natural question is whether more general arrangements of wedge sums---possibly with several summands---be consistent with an M-theoretic \ac{UV}-completion; naively, it would appear that most of these configurations be either equivalent to a perturbative string limit, intrinsically strongly coupled, or inconsistent altogether. Shedding more light on these issues, and connecting them to a broader mathematically and physically motivated framework, would be of paramount importance to progress toward a more complete understanding of quantum geometry, dualities and quantum gravity as a whole.

\section*{Acknowledgements}

It is a pleasure to thank Roberta Angius, Ilka Brunner, Robert Helling, Giorgio Leone, Joaquin Masias, Carmine Montella, Chuying Wang and Matteo Zatti for insightful discussions. The authors are grateful to Ralph Blumenhagen and Salvatore Raucci for feedback on the manuscript. The work of D.L. is supported by the German-Israel-Project (DIP) on Holography and the Swampland.

\begin{acronym}

\tooltipacro{QFT}{quantum field theory}
\acrodefplural{QFT}{quantum field theories}
\tooltipacro{CFT}{conformal field theory}
\acrodefplural{CFT}{conformal field theories}
\tooltipacro{EFT}{effective field theory}
\acrodefplural{EFT}{effective field theories}
\tooltipacro{UV}{ultraviolet}
\tooltipacro{IR}{infrared}
\tooltipacro{RNS}{Ramond-Neveu-Schwarz}
\tooltipacro{RR}{Ramond-Ramond}
\tooltipacro{NLSM}{non-linear sigma model}
\acrodefplural{NLSM}{non-linear sigma models}
\tooltipacro{GLSM}{gauged linear sigma model}
\acrodefplural{GLSM}{gauged linear sigma models}
\tooltipacro{CRP}{connected resolution property}
\tooltipacro{DRP}{disconnected resolution property}
\tooltipacro{TMF}{topological modular forms}
\tooltipacro{KK}{Kaluza-Klein}
\tooltipacro{ESC}{emergent string conjecture}
\tooltipacro{VEV}{vacuum expectation value}
\acrodefplural{VEV}{vacuum expectation values}
\end{acronym}

\bibliographystyle{ytphys}
\baselineskip=.95\baselineskip
\bibliography{ref}

\end{document}